\documentclass[twocolumn]{aastex61}

\usepackage{amsmath}
\usepackage{lipsum}
\usepackage{longtable}
\usepackage{multirow}
\shorttitle{THE ASTROPHYSICAL JOURNAL}
\shortauthors{Gaurav Singh et. al.}

\begin{document}

\title{Peculiarities in the horizontal branch stars of globular cluster NGC 1851: Discovery of a Blue straggler companion to an EHB star}

\correspondingauthor{Gaurav Singh}
\email{gaurav@aries.res.in}

%%\author[0000-0002-0786-7307]{Gaurav Singh}
\author{Gaurav Singh}
\affil{Aryabhatta Research Institute of Observational Sciences (ARIES), Manora Peak, Nainital, 263001}
\affil{Department of Physics and Astrophysics, University of Delhi, Delhi 110007}
\author{Snehalata Sahu}
\affiliation{Indian Institute of Astrophysics, Koramangala II Block, Bangalore-560034, India}
\author{Annapurni Subramaniam}
\affiliation{Indian Institute of Astrophysics, Koramangala II Block, Bangalore-560034, India}
\author{R. K. S. Yadav}
\affiliation{Aryabhatta Research Institute of Observational Sciences (ARIES), Manora Peak, Nainital, 263001}
\begin{abstract}

We present a study of Far Ultraviolet (FUV) bright horizontal branch (HB) stars to understand the peculiarities seen in the HB sequence of the Globular Cluster NGC 1851, using ground and space based multi-wavelength data. Optical and UV color magnitude diagrams are used to classify HB stars and their membership from HST and Gaia DR2 data. The Spectral energy distributions (SEDs) of the hot HB stars located from the core to tidal radii are constructed. The SEDs reveal that the HB stars near the G-jump show a decrease in the FUV flux when atmospheric models of cluster metallicity are used for fitting, but a better fit is found with higher metallicity models, as expected due to atmospheric diffusion. We report on four particularly interesting extreme HB (EHB) stars, two each in inner and outer regions. We detect a sub-luminous EHB and a ``blue-Hook'' candidates with temperatures $T_{eff}$ $\sim$ 25,000 K and 31,000 K, respectively. We found an EHB star ($T_{eff}$ $\sim$ 17,000 K) with the radius lies between the BHB and normal EHB stars. The most peculiar of our EHB stars ($T_{eff}$ $\sim$ 28,000 K) is found to be a photometric binary to a Blue Straggler star (BSS) ($T_{eff}$ $\sim$ 7,000 K), which is an important target for spectroscopic study. This discovery of the candidate EHB+BSS binary system, could help to explain the mass loss in the RGB phase, leading to the formation of EHB stars.

\end{abstract}

\keywords{Globular star clusters(656); Blue straggler stars(168); Horizontal branch stars(746); Hertzsprung Russell diagram(725)}

\section{INTRODUCTION} \label{sec:intro}

Globular clusters (GCs) are among the oldest stellar systems known in our Galaxy, and thus they provide a unique opportunity to study the late stages of stellar evolution \citep{2000ApJ...530..352D}. GCs host various stellar populations i.e., Horizontal branch (HB), Post Asymptotic giant branch (PAGB), and Blue straggler stars (BSSs), which are bright in UV. The HB stars are defined from their locus in the theoretical luminosity temperature relationship as these are core helium-burning stars. The UV wavelength is found to be an excellent probe to study the HB stars, particularly the hot end of the HB population and their characteristics \citep{1998ESASP.413..515R,1999MmSAI..70..599F,2014AJ....148..131S,2017MNRAS.469..267D}.

There are several discontinuities identified in the HB sequence of GCs. The appearance of these discontinuities depends on the applied passbands \citep{2016ApJ...822...44B}. \cite{1998ApJ...500..311F} first identified the location of gaps, which falls at a similar position in color-magnitude diagrams (CMDs) for a set of three GCs. In terms of temperatures, these discontinuities i.e., the ``Grundahl jump'' (G-jump) is located within the blue HB (BHB) sequence at $\sim$ 11,500 K \citep{1998ApJ...500L.179G,1999ApJ...524..242G}, the ``Momany jump'' (M-jump) within the extreme HB (EHB) sequence at $\sim$ 20,000 K \citep{2002ApJ...576L..65M,2004A&A...420..605M} and the gap between the EHB and ``blue-Hook'' stars, spanning $\sim$ 32,000-36,000 K (\cite{2001ApJ...562..368B,1997ApJ...474L..23S}). These features appear to be ubiquitous in the GCs having broad distribution in the temperature of HB stars. 

Stars that are hotter than the G-jump located in the BHB exhibit enhanced metal abundances through the process of atomic diffusion, i.e., through radiative levitation $\&$ gravitational sedimentation \citep{Brown_2017,1999A&A...346L...1M,2000A&A...360..120M,2006A&A...452..493P,2003ApJS..149...67B}. These BHB stars have lower surface gravities than expected from canonical HB models \citep{2001PASP..113.1162M}. The stellar evolution models developed by \cite{2007ApJ...670.1178M}, which includes atomic diffusion, have shown good agreement with the observed abundance patterns found in BHB stars. \cite{2002HiA....12..292S} and \cite{2013osp..book.....C} found that the onset of radiative levitation and the appearance of G-jump in BHB stars, may be due to disappearance of surface convection from He I ionization zone at temperatures hotter than $\sim$ 12,000 K. However, they also suggested other processes, like rotation and turbulence, as the surface convection could not be the only factor responsible for suppressing atomic diffusion in stars redder than G-jump. In the context of M-jump, \cite{Brown_2017} found that this feature ($\sim$ 18,000 K) is mainly due to the changes in the atmospheric Fe abundance. To study the observed discontinuities in the HB sequence, the UV observations of these stars play a very important role, since optical colors become degenerate at the temperatures of EHB stars \citep{2016ApJ...822...44B,2001ApJ...562..368B,2010ApJ...718.1332B,2000ApJ...530..352D,2008ApJ...677.1069D,2009MNRAS.394L..56D}. UV observations of GCs have also revealed another interesting set of stellar populations, i.e., subluminous EHB stars. They were first observed in massive GC $\omega$ Cen by \cite{1996ApJ...466..359D} $\&$ \cite{2000ApJ...530..352D} . Their evolution cannot be explained using canonical stellar evolutionary theory. They are proposed to have formed via a delayed helium-core flash due to high mass loss during RGB phase \citep{1996ApJ...466..359D}. The surface chemical composition changes induced by He-flash mixing modify the star's emergent spectral energy distribution (SED), and therefore these objects appear as sub-luminous objects in FUV CMD.

BSSs, on the other hand are defined from their location in the optical CMD, where it appears as an extrapolation of the main sequence (MS). They were first identified by \cite{1953AJ.....58...61S} in the outer region of the GC M3. Several formation mechanisms of BSSs are proposed to date, but none of them can thoroughly explain the observed features obtained in GC. At present, the main leading scenarios, are mass transfer between primordial binary companions \citep{1964MNRAS.128..147M,1976ApJ...209..734Z} and the merger of stars induced by collisions \citep{1976ApL....17...87H,1989AJ.....98..217L}. Recently, a new formation channel proposed for twin BSSs formed in compact binary system \citep{Zwart_2019}, which involves mass transfer via a circumbinary disk from an evolved outer tertiary companion.  

To study the particular UV-bright (EHB, blue-Hook, and BSS) stars, UVIT/AstroSat observations are very useful. UVIT/AstroSat study of the GCs NGC 1851 and NGC 288 by \cite{2017AJ....154..233S} and \cite{2019MNRAS.482.1080S}, respectively, have shown the importance of UVIT/AstroSat in understanding the HB morphology of the GCs. UVIT/AstroSat observations have also been beneficial in detecting hot companions to BSSs. White Dwarf (WD) companions to BSSs in M67 \citep{2019ApJ...886...13J,2019ApJ...882...43S} and a PAGB/HB companion to a BSS in NGC 188 \cite{2016ApJ...833L..27S} were found using UVIT observations in multiple filters and supported by SED analysis. Also, \cite{2019ApJ...876...34S}, detected a WD companion to a BSS in the outskirts of the GC NGC 5466.

In this paper, we present the study of FUV bright HB stars in the GC NGC 1851 from core ($r_{c}$ $\sim$ $5\farcs4$) to the tidal radius ($r_{t}$ $\sim$ $6\farcm5$). NGC 1851 ($\alpha_{J2000}$ = $5^{h}$ $14^{m}$ $6^{s}.76$, $\delta_{J2000}$ = $-40\degr$ $2\arcmin$ $47\farcs6$; $l$ = $244.\degr$51, $b$ = $-35.\degr$03) \footnote{\url{http://physwww.physics.mcmaster.ca/~harris/mwGC.dat}} \citep{1996AJ....112.1487H} is located at a distance of 12.1 kpc from the Sun. This is a high density globular cluster and has an intermediate metallicity ([Fe/H] = $-$1.18). \cite{2017AJ....154..233S} presented the first UV CMDs of the cluster NGC 1851 using UVIT/AstroSat telescope, whereas the membership estimations of the populations were not available. Our aim is to provide a proper motion cleaned UV and Optical CMDs covering the entire cluster region and derive the properties of HB stars in the cluster.

The paper is arranged as follows: Sections 2 outlines the data used and reduction, the membership probability estimation in Section 3, UV CMDs, and the results obtained through SED analysis in Section 4, followed by discussion, summary, and conclusions in Sections 5 and 6.

\section{DATA USED AND REDUCTION} \label{sec:style}
\subsection{The data sets} \label{subsec:data}
\begin{figure}[ht!]
\epsscale{1.27}
\plotone{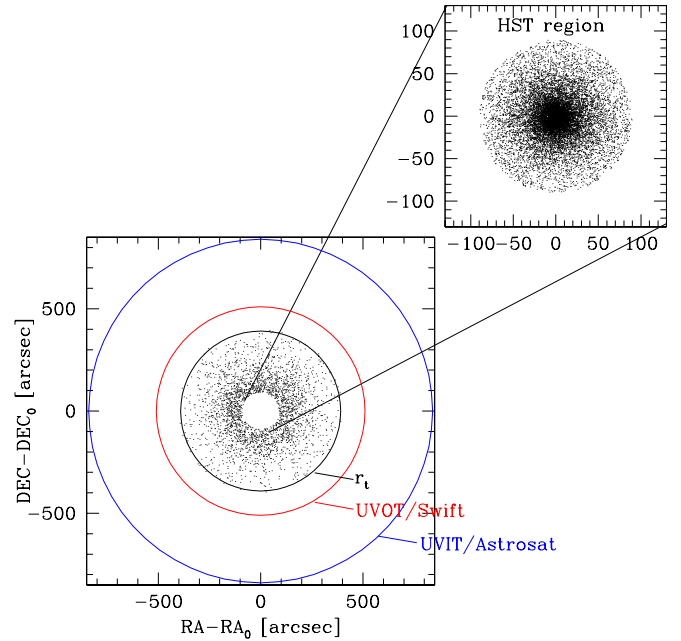}
\caption{The map showing the distribution of stars in the entire cluster region. The map shown in the top right panel contains the central crowded HST region common to both the WFC3 and ACS FOV. The map in the center-left is plotted for the stars common to UVIT/AstroSat, UVOT/Swift, Stetson's photometry, and GAIA DR2, and it extends from inner HST region up to the tidal radius. 
\label{fig:dist}}
\end{figure}

To study the FUV bright HB stars in NGC 1851, we use the multi-wavelength space and ground based data sets to cover the entire cluster extension ($r$ $\sim$ $r_{t}$), as shown in Figure \ref{fig:dist}.

In the dense central region of the cluster ($r$ $\leq$ $80\arcsec$), we use the HST data in the astro-photometric catalog \cite{2018MNRAS.481.3382N}, %as
a part of HST UV Legacy Survey of Galactic Globular Cluster. In this catalog, the data set in \textit{F275W}, \textit{F336W} and \textit{F438W} filters were observed through WFC3/UVIS channel, while the data set in \textit{F606W} and \textit{F814W} filters %come 
come from ACS/WFC channel \citep{2015AJ....149...91P}. The catalog also contains the membership information of all the stars common to WFC3/UVIS and ACS/WFC field of view (FOV), which defines the circular HST region considered for the present analysis. 
The Far-UV (FUV) and Near-UV (NUV) data used for the analysis are obtained from the UVIT instrument on-board the \textit{AstroSat} satellite. UVIT/AstroSat is primarily an imaging instrument that consists of twin 38 cm telescopes. It provides imaging in FUV channel (130-180 nm) and NUV channel (200-300 nm) simultaneously with a circular field of view $\sim$ $28\arcmin$. The angular resolution of UVIT/AstroSat is $<1\farcs8$ in the FUV and NUV channels. The details of instrument and calibration are described in \cite{2016SPIE.9905E..1FS,2017AJ....154..128T}. For this cluster, %we did 
the observations were performed in three filters: $\textit{F148W}$, $\textit{F169M}$, and $\textit{N279N}$, as a part of Performance Verification (PV) phase during $19^{th}$-$21^{st}$ March 2016. The details of data acquisition and reduction procedures are provided in \cite{2017AJ....154..233S}. 

Apart from the availability of N279N filter from the UVIT/AstroSat, we need more data sets in the NUV region to fit the UV part of the spectra of HB stars. Therefore, for the complimentary UV data in the outer region ($80\arcsec \leq r \leq r_{t}$), we used the Ultraviolet/Optical Telescope (UVOT) data. UVOT/Swift is a modified Richey-Chretien telescope on-board Swift spacecraft. It provides a FOV of about 17$'\times 17' $ with an angular resolution of $2\farcs3$. UVOT/Swift consists of broadband ultraviolet ($\textit{uvw1}$, $\textit{uvm2}$, and $\textit{uvw2}$) filters \citep{2005SSRv..120...95R,10.1111/j.1365-2966.2007.12563.x}. UVOT/swift data reduction of the cluster is described in the following subsection.

The detailed log of observations from UVIT/AstroSat and UVOT/Swift with the information of filters and exposure times are listed in Table \ref{tab:dec}.

\begin{table}[h!]
\renewcommand{\thetable}{\arabic{table}}
\centering
\caption{Description of the UVIT/AstroSat and UVOT/Swift data sets, with the details of central wavelength, FWHM and exposure times, respectively.} 
\begin{tabular}{ccccc}
\tablewidth{0pt}
\hline
\hline
Filter & Central Wavelength & FWHM & Exp Time\\
& \AA & \AA & Ks
\\
\hline
\multicolumn{4}{c}{UVIT/AstroSat}\\\hline
%%\decimals
$F148W$ & 1481 & 500 & 6.98\\
$F169M$ & 1608 & 290 & 5.27\\
$N279N$ & 2792 & 90 & 12.23\\
\hline
\multicolumn{4}{c}{UVOT/Swift}\\\hline
$uvw1$ & 2600 & 693 & 1.25\\
$uvm2$ & 2246 & 498 & 4.31\\
$uvw2$ & 1928 & 657 & 2.15\\
\hline
\hline
\label{tab:dec}
\end{tabular}
\end{table}

To cover the optical wavelength in the outer region, we used the Stetson's $\textit{UBVRI}$ photometric catalog \citep{2019MNRAS.485.3042S}. We used Gaia DR2 data to compute the membership of stars located in the regions outside the HST region. The procedure of membership determination is described in the following subsection.

\subsection{The data reduction and calibration of UVOT/Swift data sets} 
We obtained the raw and pipeline processed data from the HEASARC archive\footnote{\url{https://heasarc.gsfc.nasa.gov/cgi-bin/W3Browse/w3browse.pl}}. These include the exposure maps, auxiliary spacecraft data, and the target's processed images, which are geometrically corrected for sky coordinates. For reducing the UVOT/Swift data, we followed the procedure described in \cite{2014AJ....148..131S}.

We generated large scale sensitivity (LSS) maps for individual frames using the UVOTSKYLSS task. These frames were combined into a single multi-extension file using the FAPPEND task. We then used UVOTIMSUM to combine and generate a single image for each filter and their corresponding sensitivity and exposure map images.

\begin{figure}[ht!]
\epsscale{1.24}
\plotone{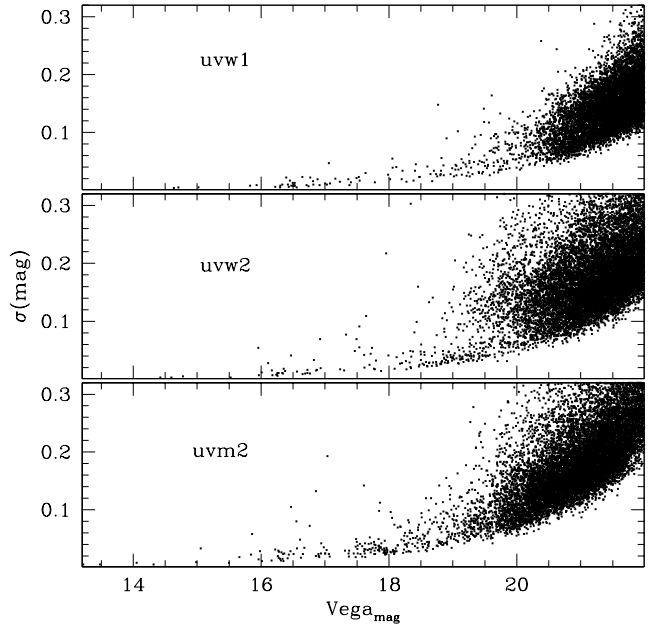}
\caption{Photometric errors as a function of magnitudes are plotted for the UVOT/Swift filters available for NGC 1851. \label{fig:errors}}
\end{figure}

We performed PSF photometry using DAOPHOT software developed by \cite{1987PASP...99..191S}. To perform PSF photometry, we selected bright and isolated stars across the FOV to obtain the PSF characteristics for the UVOT/Swift images. The PSF fit errors in photometry as a function of magnitudes are plotted in Figure \ref{fig:errors}. To determine the aperture correction, we also performed aperture photometry using aperture radius of $5\farcs0$. 

To calibrate the instrumental magnitudes, we first applied the aperture correction to PSF magnitudes. The aperture corrected magnitudes are converted to count rates. Background subtraction on the count rates was performed using the UVOTSOURCE task for an aperture of $5\farcs0$ in the background region of the combined UVOT/Swift images.

We corrected the raw count rates obtained after background subtraction to the coincidence loss and the LSS using the LSS map. Finally, we converted the corrected count rates to the VEGA MAG photometry system using the zero-point corrections taken from \cite{2010MNRAS.406.1687B}. 

\section{Membership probability and selection of members} \label{sec:conta}
To select the genuine members of the cluster for further analysis, we used the membership probability information given in the astro-photometric catalog of \cite{2018MNRAS.481.3382N} for the HST region. We have used \textit{Gaia} DR2 proper motion data to obtain the membership probability of stars in the outer region. Thus, we have the membership probability for stars in the entire cluster region ($r$ $\sim$ $r_{t}$), which is used for the selection of genuine cluster members \citep{10.1093/mnras/sty2961}. 

The \textit{Gaia} DR2 catalog provides the photometric and astrometric information for all the stars up to $G$ $\sim$ 21 mag \citep{2016A&A...595A...1G,2018A&A...616A...1G}. To derive the membership information of all the stars in the outer region, we adopted the method described in \cite{1971A&A....14..226S} based on the maximum likelihood principle. 
\begin{figure}[ht!]
\epsscale{1.26}
\plotone{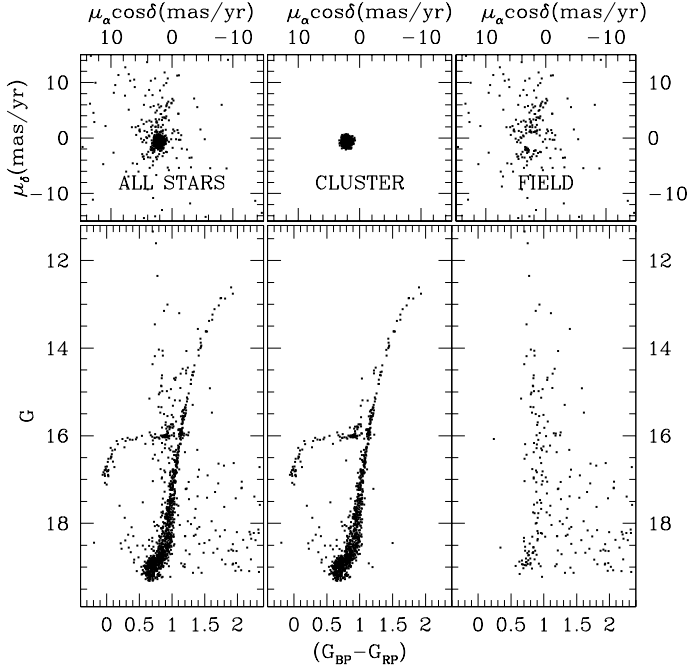}
\caption{The VPDs and their corresponding CMDs of all stars, cluster members, and field stars are plotted (from left to right). The criteria for membership estimation is shown by a circle of radius 1.5 mas/yr. We have considered only the stars with good astrometric measurements i.e., with nper $\textgreater$ 8 and nal $\textgreater$ 120 for membership estimation.\label{fig:vpd}}
\end{figure}

The vector point diagrams (VPDs) and corresponding $\textit{Gaia}$ CMDs for all the stars lying in the outer region are shown in Figure \ref{fig:vpd}. By considering the values estimated from \cite{2018A&A...616A..12G} as the initial guess, we have taken the centers of the cluster distribution at ($2.13$, $-0.62$) mas/yr. As shown in the middle panel, the membership selection criteria of radius 1.5 mas/yr in the VPD, is used for obtaining the distribution function for cluster and field members.

To derive the membership probability, we estimate the frequency distribution of cluster stars ($\phi^{\nu}_{c}$) and field stars ($\phi^{\nu}_{f}$). The frequency distribution function for the $i^{th}$ star of a cluster and the field can be written as follows:

\begin{multline}
\phi^{\nu}_{c} = \frac{1}{{2\pi}\surd{(\sigma^{2}_{c}+ \epsilon^{2}_{xi})(\sigma^{2}_{c}+ \epsilon^{2}_{yi})}} \\
\times exp \left \{ -\frac{1}{2} \Bigg[ \frac{(\mu_{xi}-\mu_{xc})^{2}}{\sigma^{2}_{c}+ \epsilon^{2}_{xi}} + \frac{(\mu_{yi}-\mu_{yc})^{2}}{\sigma^{2}_{c}+ \epsilon^{2}_{yi}}
\Bigg] \right \} 
\end{multline}

\begin{multline}
\phi^{\nu}_{f} = \frac{1}{{2\pi}\surd{(1-\gamma^{2})}\surd{(\sigma^{2}_{xf}+ \epsilon^{2}_{xi})(\sigma^{2}_{yf}+ \epsilon^{2}_{yi})}} exp \Bigg \{ -\frac{1}{2(1-\gamma^{2})}\\ 
\Bigg[ \frac{(\mu_{xi}-\mu_{xf})^{2}}{\sigma^{2}_{xf}+ \epsilon^{2}_{xi}} - \frac{2\gamma(\mu_{xi}-\mu_{xf})(\mu_{yi}-\mu_{yf})}{\surd{(\sigma^{2}_{xf}+ \epsilon^{2}_{xi})(\sigma^{2}_{yf}+ \epsilon^{2}_{yi})}} + \frac{(\mu_{yi}-\mu_{yf})^{2}}{\sigma^{2}_{yf}+ \epsilon^{2}_{yi}}
\Bigg] \Bigg \} 
\end{multline}

where $\mu_{xi}$ and $\mu_{yi}$ represent the proper motions of the $i^{th}$ star. The $\mu_{xc}$ and $\mu_{yc}$ denotes the cluster's proper motion center, $\mu_{xf}$ and $\mu_{yf}$ denotes the field proper motion center, $\epsilon_{xi}$ and $\epsilon_{yi}$ are the observed errors in proper motion components, $\gamma$ is the correlation coefficient, $\sigma_{c}$ is the cluster intrinsic proper motion dispersion, while $\sigma_{xf}$ and $\sigma_{yf}$ represents the field intrinsic proper motion dispersion.

The value of $\gamma$ can be calculated as

\begin{gather}
\gamma = \frac{(\mu_{xi}-\mu_{xf})(\mu_{yi}-\mu_{yf})}{\sigma_{xf}\sigma_{yf}}
\end{gather}

\begin{figure}[ht!]
\epsscale{1.24}
\plotone{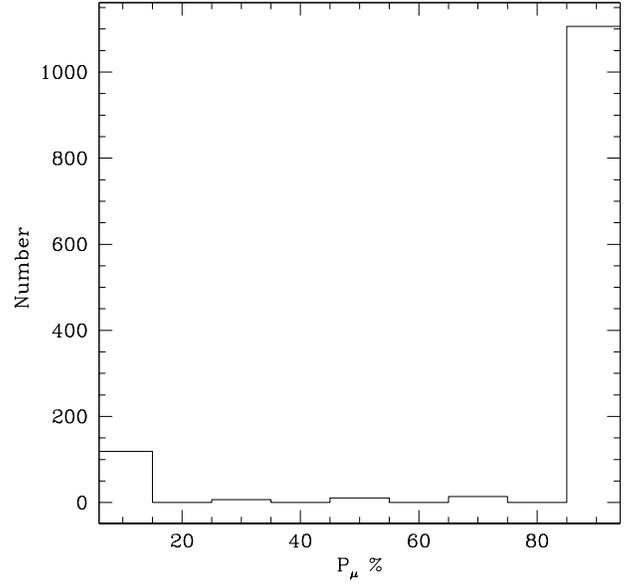}
\caption{The membership estimation for the stars lying in the outer region is shown in the histogram plot. The plot shows that most of the stars have membership probability $\textit{P}$ $\textgreater$ 80 $\%$. \label{fig:mem}}
\end{figure}

To obtain the distribution function $\phi^{\nu}_{c}$ and $\phi^{\nu}_{f}$, we considered the stars with error in proper motion better than 1 mas/yr. For, good astrometric measurements, we considered stars with nper $\textgreater$ 8 and nal $\textgreater$ 120, as described in \cite{Kim_2019}, where, ``nper'' is the number of visibility periods used in the Astrometric solution and ``nal'' is the total number of observations. We obtain the mean and standard deviations of the cluster distribution to be $\mu_{xc}$ = 2.134 mas/yr and $\mu_{yc}$ = $-0.612$ mas/yr, with $\sigma_{c}$ = 0.35 mas/yr. For field stars, we obtained the centers to be $\mu_{xf}$ = 3.25 mas/yr , $\mu_{yf}$ = 1.13 mas/yr, $\sigma_{xf}$ = 6.15 mas/yr and $\sigma_{yf}$ = 8.48 mas/yr. 

To obtain the distribution of stars, we use 

\begin{gather}
\phi = n_{c} \phi^{\nu}_{c} + n_{f} \phi^{\nu}_{f}
\end{gather}

where $n_{c}$ and $n_{f}$ are the normalized number of stars found for cluster
and field ($n_{c}$ + $n_{f}$ = 1). Hence, the membership probability
for $i_{th}$ star is given by

\begin{gather}
P^{\mu} (i) = \frac{\phi_{c} (i)}{\phi (i)}
\end{gather}

Figure \ref{fig:mem} shows the histogram of membership probability estimated using the Gaia DR2 catalog for all the stars lying in the outer region is plotted. The histogram shows that most of the stars have membership probability $\textit{P}$ $\textgreater$ 80 $\%$. The total number of stars in the outer region having membership probability $\textit{P}$ $\textgreater$ 80 $\%$ and with proper astrometric measurements was found to be $\sim$ 1,106.  

We combined the HST and Gaia data for obtaining the membership catalog of all the stars in the entire cluster region. For further analysis, we have considered only stars with a membership probability $\textit{P}$ $\textgreater$ 80 $\%$ as genuine members for the entire cluster region.\\

\section{Analysis and Results} \label{sec:results}
To select the FUV bright HB stars located in the inner region ($r_{c}$ $\leq$ $r$ $\leq$ $80\arcsec$), we have cross-matched the UVIT/AstroSat data with HST photometry data. We have excluded the dense core region ($r$ $\leq$ $r_{c}$) of the cluster to avoid multiple cross-match in the HST region. To obtain a catalog of FUV bright HB stars in the outer region ($80\arcsec$ $\textless$ $r$ $\leq$ $r_{t}$), we have cross-matched the UVIT/AstroSat photometry with the UVOT/Swift photometry along with XMM/OM \citep{2012MNRAS.426..903P} and GALEX data \citep{2011Ap&SS.335..161B} within $2\arcsec$. We then cross-matched the UV catalog with the Stetson's $UVBRI$ photometry, {\it C}, {\it T1} and {\it T2} filters from the Washington telescope \citep{2014AJ....148...27C}, and {\it J}, {\it K} photometry by \cite{2015AJ....150..176C} within $2\arcsec$, to get a final catalog of FUV bright HB stars, extending from UV to IR wavelengths for both the inner and outer region. In this way, we have a full sample of HB stars extending from the core radius ($r_{c}$) to the tidal radius ($r_{t}$) of the cluster. The main aim of the present study is to estimate the FUV bright HB stars' parameters by fitting their SEDs with theoretical spectra.

\subsection{UV CMD of NGC 1851 in the inner region}
\begin{figure}[ht!]
\epsscale{1.27}
\plotone{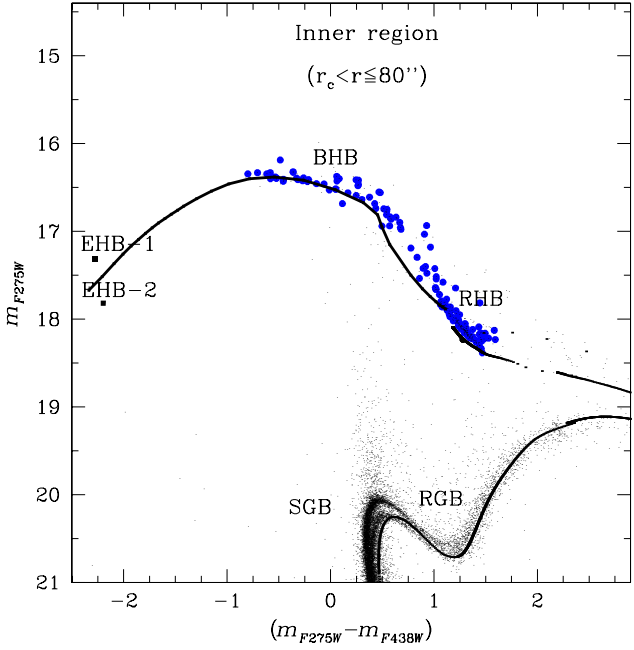}
\caption{The UV CMD ($m_{F275W}$, ($m_{F275W}$ - $m_{F438W}$)) of the cluster NGC 1851 is plotted. The HB stars used in our analysis are shown with blue dots. These stars are located in the inner region ($r_{c}$ $\leq$ $r$ $\leq$ $80\arcsec$).\label{fig:hst}}
\end{figure}

We have plotted the UV-optical color-magnitude diagram (CMD) ($m_{F275W}$, ($m_{F275W}$ - $m_{F438W}$)) of the cluster NGC 1851 using HST filters, as shown in Figure \ref{fig:hst}. We found a sample of the RHB and BHB population along with two EHB stars in the inner region. The stars selected from the inner region ($r_{c}$ $\leq$ $r$ $\leq$ $80\arcsec$) shown with blue dots will be used in further analysis. We have avoided the central core region due to the large crowding of the stars. 

The UV CMD is fitted with the HB model generated using the updated Bag of Stellar Tracks and Isochrones (BaSTI-IAC\footnote{\url{http://basti-iac.oa-abruzzo.inaf.it/hbmodels.html}}, \cite{Hidalgo_2018} ). The BaSTI-IAC isochrone shown in Figures 5-8 is generated for a metallicity of [Fe/H] = $-1.2$ dex \citep{2013AJ....145...25K}, a distance modulus of 15.47 mag and an assumed age of 10 Gyr \citep{2008ApJ...672L.115C}. The HB model includes the ZAHB and POST-ZAHB evolutionary tracks with the core mass ranging from 0.492 $M_{\odot}$ to 0.800 $M_{\odot}$. The HB model includes diffusion and is generated for a metallicity of [Fe/H] = $-1.2$ dex.

\subsection{Morphology and selection of HB stars in the inner region} \label{sec:morph}
To classify the HB stars into various subgroups within HB, we used the color-color plane (CCP) ($C_{F275W, F336W, F438W}$ versus ($m_{F275W} - m_{F438W}$)) defined by \cite{2013ApJ...767..120M}, where $C_{F275W, F336W, F438W}$ = ($m_{F275W} - m_{F336W}$) $-$ ($m_{F336W} - m_{F438W}$). We prefer to use CCP as a primary selection criteria to understand the peculiarities in the HB sequence, since the discontinuities present in the HB morphology get amplified in $C_{F275W, F336W, F438W}$ index \citep{2016ApJ...822...44B}. 

\begin{figure}[ht!]
\epsscale{1.26}
\plotone{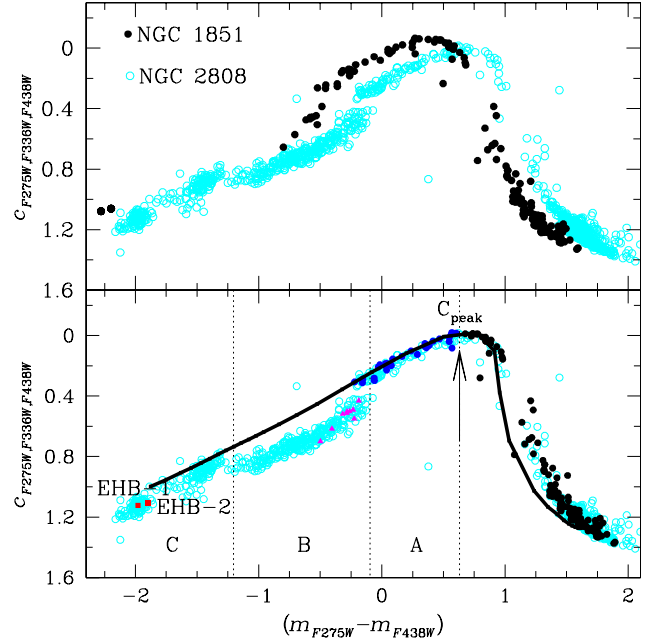}
\caption{The color-color plane (CCP) of NGC 1851 and the cluster NGC 2808 are plotted in the $upper$ $panel$. The $lower$ $panel$ shows the color-color plane of NGC 1851 is over-plotted on CCP of the HB stars of the cluster NGC 2808 by applying a shift of (0.30, 0.045). The positions of $C_{peak}$, G-jump and M-jump are located at $8,500$ K, $11,500$ K, and $20,000$ K respectively. The stars located in Region A, B and C are marked with blue circles, pink triangles and red square, respectively. The CCP is fitted with the BASTI-IAC model, show deviation for the stars located in the Region B.\label{fig:ccp}}
\end{figure}

To understand the HB morphology of NGC 1851, we have taken HB stars of NGC 2808 as a reference as considered by \cite{2016ApJ...822...44B}. NGC 2808 is taken as a reference because it is known to have a long HB sequence \citep{1997ApJ...480L..35S,2000A&A...363..159B}. We plotted the HB stars of the cluster NGC 1851 and NGC 2808 with black and cyan colors, respectively, in the $upper$ $panel$ of Figure \ref{fig:ccp}. We used $C_{peak}$ as a reference point to match the CCP of the two clusters as described in \cite{2016ApJ...822...44B}. The corresponding shift in the $C_{peak}$ of the two clusters was found to be (0.30, 0.045) along the two axes of the CCP. Therefore, we applied this shift to overplot the CCP of NGC 1851 on the CCP of the cluster NGC 2808. 
We marked the positions of discontinuities in the CCP as follows: the G-jump ($\sim$ 11,500 K), the M-jump ($\sim$ 20,000 K), along with $C_{peak}$ ($\sim$ 8,600 K) \citep{2016ApJ...822...44B} and shown with dotted lines in the $lower$ $panel$ of Figure \ref{fig:ccp}.

To understand the UV emission properties of the HB stars around these discontinuities, we divided the CCP into three regions based on their temperatures; 
\begin{enumerate} 

\item Region A: This includes the HB stars located between $C_{peak}$ and G-jump.
\item Region B: In this, we have selected the HB stars located between G-jump and M-jump.

\item Region C: This includes the stars that are located after M-jump. 

\end{enumerate}

The HB stars selected from these three regions A, B, C from CCP, as shown in the $lower$ $panel$ of Figure \ref{fig:ccp}, are marked as blue circles, pink triangles, and red square, respectively.

Going through the definition of \cite{2016ApJ...822...44B}, we define the stars hotter than the RR Lyrae instability strip ($\sim$ 8,000 K) and cooler than $\sim$ 20,000 K as BHB stars and stars that are hotter than 20,000 K are EHB stars. 

In the $lower$ $panel$ of Figure \ref{fig:ccp}, we have fitted the BASTI-IAC model \citep{Hidalgo_2018}.

There is an observed deviation from canonical HB model in $C_{F275W, F336W, F438W}$ index in the color range $-1.2$ $\textless$ ($m_{F275W}$-$m_{F438W}$) $\textless$ $-0.2$. Stars located near ($m_{F275W} - m_{F438W}$) $\geq$ $-1.8$ mag, define the canonical end of the EHB stars. We can also classify them as blue-Hook stars if they fall within 1.0 $\textless$ $C_{F275W, F336W, F438W}$ $\textless$ 1.3 mag and $-2.4$ $\textless$ ($m_{F275W} - m_{F438W}$) $\textless$ $-1.9$ mag in the CCP. 

In Region A, since BHB stars are cooler than $\sim$ 11,500 K, they are not subjected to atomic diffusion. The BHB stars located in Region B, are basically stars which have undergone atomic diffusion, while the stars located in the Region C are EHB stars or blue-hook stars depending upon their location around the gap region. We will use this classification to understand the UV properties of these HB stars and the fitting parameters obtained by fitting their corresponding SEDs.

\subsection{UV-Optical CMDs of the FUV bright HB stars} \label{sec:floats}
To select the FUV bright HB stars from both the inner ($r_{c}$ $\leq$ $r$ $\leq$ $80\arcsec$) and outer region ($80\arcsec$ $\textless$ $r$ $\leq$ $r_{t}$), we have considered the UVIT/AstroSat filters, which is common to both the region.

Hence we created UV and UV-optical CMDs of the FUV bright HB stars located in both the regions. We have transformed the F606W magnitude to Johnson $V$ magnitude available in the Stetson's catalog, using the relationship given by \cite{2005PASP..117.1049S}. 

\begin{figure}[ht!]
\epsscale{1.25}
\plotone{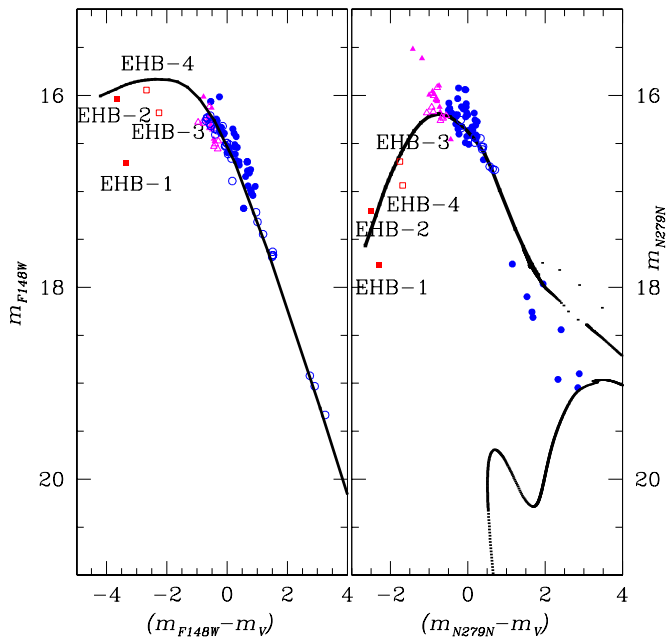}
\caption{The UV CMD plots ($m_{FUV}$, ($m_{FUV}$ - $m_{V}$) $\&$ $m_{NUV}$, ($m_{NUV}$ - $m_{V}$)) of FUV bright HB stars located in both the inner and outer region are shown by filled and open symbols, respectively. The BASTI-IAC fits well the hot BHB sequence, where diffusion and mass loss are more prominent. However, EHB-1 and EHB-1 show significant deviation from the BAST-IAC model. Please see the location of EHB-4 star in Figure \ref{fig:bss}. \label{fig:whole}}
\end{figure} 

\begin{figure}[ht!]
\epsscale{1.25}
\plotone{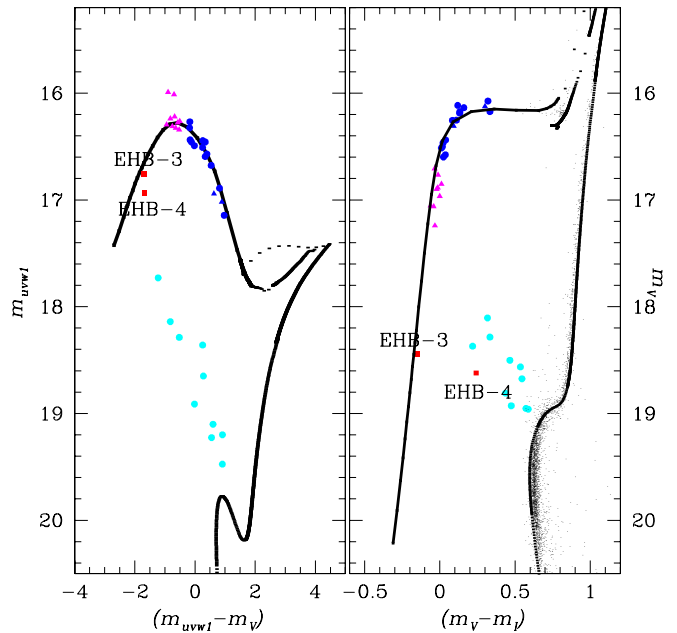}
\caption{The UV-Optical ($m_{uvw1}$ - $m_{V}$) and Optical ($m_{V}$ - $m_{I}$) CMD of the FUV bright HB stars located in the outer region. Few of the BHBd stars are bright in the UV-Optical CMD plotted in the $left$ $panel$, and they are also relatively bright in FUV and NUV, as can be seen in UV CMDs of Figure 7 as well. The location of EHB-4 star in the optical CMD lies in the BSS sequence (shown with cyan color) in the $right$ $panel$, while in the UV CMD, the location of this star shifts towards the hot end of the HB sequence in the $left$ $panel$.\label{fig:bss}}
\end{figure}

\begin{figure*}[ht!]
\centering
\epsscale{1.0}
\plotone{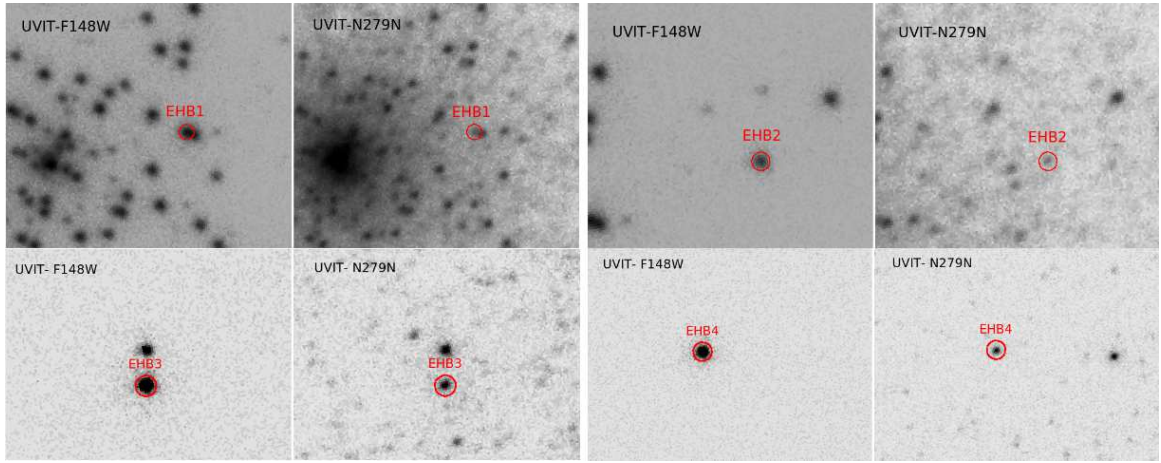}
\caption{The location of EHB population are plotted in the FUV (F148W) and NUV (N279N) images of the UVIT/AstroSat, along the left and right panels respectively.}
\label{fig:space}
\end{figure*}

In Figure \ref{fig:whole}, we plot the FUV-Optical ($m_{F148W}$, ($m_{F148W}$ - $m_{V}$)) and NUV-Optical ($m_{N279N}$, ($m_{N279N}$ - $m_{V}$)) CMDs of FUV bright HB stars found in both the regions. Stars that are selected from the inner region using CCP are plotted in the CMDs along with HB stars located in the outer region, which are shown as filled and open symbols, respectively. 

Stars that are located after the G-jump and are found in the region B of the CCP, are shown as pink triangles in the NUV-optical CMD, these are mentioned as BHBd stars for the rest of this paper. Also, the stars located following the canonical BHB sequence are shown by blue dots. The stars located in the hot end of the HB sequence are shown by red squares.
We have plotted the theoretical isochrones, which are generated using the BaSTI-IAC models \citep{Hidalgo_2018}.

We plotted the NUV-optical ($m_{uvw1}$, ($m_{uvw1} - m_{V}$)) CMD and Optical ($m_{V}$, ($m_{V}$ - $m_{I}$)) CMD for the FUV bright stars in the outer region in Figure \ref{fig:bss}. We also plotted the BSSs found in the outer region to understand their position with respect to the EHB stars in the respective CMDs.

In total, we found 71 BHB, 22 BHBd stars from both the inner and outer regions. The BHB and BHBd stars found in our study matches with the study previously done by \cite{2017AJ....154..233S}. We also detected four EHB stars that were not reported previously. The reason for not detecting these EHB stars previously was the selection criteria used by \cite{2017AJ....154..233S}, to minimize the field star contamination. Thanks to $\textit{Gaia}$ proper motion information, we can now confirm their membership. We have also checked for cross-match of these FUV sources in the inner region of the cluster. The location of four EHB stars in FUV (F148W) and NUV (N279N) images of UVIT/AstroSat are plotted in Figure \ref{fig:space}. The EHB-1 star is located at $r$ $\sim$ 1.1 $r_{h}$ from the center. 

Another FUV bright star is found adjacent to EHB-1. As we have performed PSF photometry, we are able to estimate the magnitude of EHB-1 with negligible contamination from the neighbouring star. However, the other EHB population are isolated FUV sources. The EHB-2 star which is located in inner region lies at $r$ $\sim$ 2.4 $r_{h}$. The EHB population (EHB-3 \& EHB-4) are located in the outer region at $r$ $\sim$ 4.9 $r_{h}$ and $\sim$ 8.1 $r_{h}$, respectively. 

\subsection{SED of FUV bright HB Stars} \label{subsec:def}
We use SED to obtain the fundamental parameters i.e., the total luminosity (L/$L_{\odot}$), the effective temperature ($T_{eff}$), and the radius (R/$R_{\odot}$) of FUV bright HB stars. We used observed photometric data to construct the SEDs. The SEDs are fitted using the Castelli/Kurucz stellar atmospheric models \citep{1997A&A...318..841C,castelli2004new}. The Castelli/Kurucz models use ATLAS9 stellar atmospheres and provides $T_{eff}$ in the range of 3500K to 50000K and log$g$ values from 0.0 to 5.0 dex. The Castelli/Kurucz models are computed for scaled solar abundances considering LTE with the enhanced $\alpha$ element abundances and for various values of metallicities ([Fe/H]) \citep{2003IAUS..210P.A20C}. The ATLAS9 code is consistent with the SYNTHE code, in terms of same input and numerical approaches used in solving these codes \citep{1997A&A...318..841C}. There are some differences in the opacity computations in the sense that SYNTHE code does not include any molecular continuous opacity while Castelli/Kurucz model considers molecular continuous opacity in their computation of stellar atmospheric models. 

The SEDs are built using a virtual observatory tool, Virtual Observatory SED analyzer (VOSA) \citep{2008A&A...492..277B}. Under the VOSA, synthetic photometric fluxes are calculated for a theoretical model for each of the filter transmission curves. To obtain the best fit parameters of the SED, VOSA conducts a $\chi^{2}$ minimization test by comparing the synthetic flux with the observed flux. The reduced $\chi^{2}$ value can be calculated using the formula given below.

\begin{multline}
\chi^{2} = \frac{1}{{N-N_{f}}} \sum \left \{ \frac{(F_{o,i}-M_{d}F_{m,i})^{2}}{\sigma^{2}_{o,i}} \right \} 
\end{multline}

where $N$ and $N_{f}$ represent the number of photometric data points and number of free parameters in the model, respectively. $F_{o,i}$ and $\sigma_{o,i}$ are the observed flux and corresponding error, while $M_{d}F_{m,i}$ is model flux of the star. $M_{d}$ ($(R/D)^{2}$) denotes the scaling factor corresponding to the star with the radius $R$ and distance $D$.\\
 
To fit the observed SED of the FUV bright HB stars, we used Castelli/Kurucz stellar atmospheric models in the wavelength range UV to IR. The free parameters of the Castelli/Kurucz models are [Fe/H], $T_{eff}$, and log$g$. To fit the SED, we varied the [Fe/H] in the range $-$2.0 to $+$0.5.

We plot the SED using UVIT/AstroSat and HST photometry data sets for the stars located in the inner region. We plot the SED of the stars in the outer region for which we get the maximum availability of data points common to UVIT/AstroSat, UVOT/Swift, XMM/OM, GALEX, $UVBRI$, {\it C}, {\it T1}, {\it T2}, and {\it J}, {\it K} photometric data points. We therefore obtained a range of data points from 7 to 20, covering the wavelength from FUV to NIR for both the inner and outer region stars.

The SED fit parameters estimated for all the FUV bright BHB and BHBd stars are listed in Table 2. The $T_{eff}$ of BHB stars varies from 7,250 K to 11,000 K. The corresponding values of $L/L_{\odot}$ and $R/R_{\odot}$ varies from 34.18 to 50.97 and 1.59 to 4.23, respectively. For, BHBd star, the $T_{eff}$ estimation varies from 11,000 K to 12,250 K and the corresponding values of $L/L_{\odot}$ and $R/R_{\odot}$ varies from 30.38 to 40.72 and 1.22 to 1.71, respectively.

\begin{longtable*}{ccccccccc}
\caption {SED fit parameters of the FUV bright BHB and BHBd stars for a varying metallicity. The values of temperatures, radii and luminosities of the stars derived from the SED fitting are listed in Column 5, 6, and 7 respectively. The number of points considered in fitting the SED i.e., $N_{fit}$, is listed in Column 9.}\label{Tab:2}\\\toprule
\endfirsthead
\caption* {\textbf{-- continued from previous page}}\\\toprule
\endhead
\endfoot
Star ID & RA (J2000) [$\degr$] & DEC (J2000) [$\degr$] & [Fe/H] & $T_{eff}$ [K] & $R$/$R_{\sun}$ & $L$/$L_{\sun}$ & $\chi^{2}_{reduced}$ & $N_{fit}$\\
\hline
%%\decimals
BHB-1 \footnote{The stars having spectroscopic determination of temperatures} & 78.478279 & -40.046459 & -0.5 & 10,000 $\pm$ 500 & 2.08 $\pm$ 0.06 & 39.09 $\pm$ 1.00 & 10.17 & 18\\
BHB-2 $^{4}$ & 78.536659 & -40.009552 & -0.5 & 10,250 $\pm$ 500 & 1.86 $\pm$ 0.05 & 34.45 $\pm$ 0.30 & 4.64 & 11\\
BHB-3 $^{4}$ & 78.504776 & -40.068920 & -1.5 & 9,500 $\pm$ 500 & 2.31 $\pm$ 0.06 & 39.17 $\pm$ 0.51 & 14.98 & 11\\
BHB-4 $^{4}$ & 78.565636 & -40.039619 & -1.0 & 9,500 $\pm$ 500 & 2.24 $\pm$ 0.06 & 36.93 $\pm$ 0.32 & 6.49 & 11\\
BHB-5 $^{4}$ & 78.554604 & -40.024544 & -0.5 & 10,250 $\pm$ 500 & 1.79 $\pm$ 0.04 & 31.86 $\pm$ 0.39 & 10.25 & 11\\
BHB-6 $^{4}$ & 78.495796 & -40.018997 & -1.5 & 8,250 $\pm$ 500 & 3.15 $\pm$ 0.10 & 41.63 $\pm$ 0.97 & 8.41 & 17\\
BHB-7 $^{4}$ & 78.524956 & -40.019096 & -1.5 & 8,250 $\pm$ 500 & 3.27 $\pm$ 0.10 & 44.60 $\pm$ 0.55 & 10.74 & 11\\
BHB-8 $^{4}$ & 78.539978 & -40.063988 & -1.5 & 9,500 $\pm$ 500 & 2.33 $\pm$ 0.02 & 39.75 $\pm$ 0.23 & 7.46 & 8\\
BHB-9 $^{4}$ & 78.503731 & -40.039501 & -0.5 & 9,000 $\pm$ 500 & 2.84 $\pm$ 0.08 & 47.74 $\pm$ 0.16 & 14.70 & 8\\
BHB-10 $^{4}$ & 78.583763 & -40.025261 & -0.5 & 10,750 $\pm$ 500 & 1.66 $\pm$ 0.04 & 33.22 $\pm$ 0.85 & 17.03 & 18\\
BHB-11 $^{4}$ & 78.571258 & -40.069687 & -0.5 & 11,000 $\pm$ 500 & 1.59 $\pm$ 0.04 & 33.56 $\pm$ 0.90 & 16.04 & 18\\
BHB-12 $^{4}$ & 78.529327 & -40.025394 & -1.5 & 11,000 $\pm$ 500 & 1.61 $\pm$ 0.04 & 34.38 $\pm$ 0.21 & 9.95 & 8\\
BHB-13 & 78.501678 & -40.061771 & -1.5 & 7,750 $\pm$ 500 & 3.68 $\pm$ 0.13 & 43.18 $\pm$ 1.28 & 19.61 & 16\\
BHB-14 & 78.433861 & -40.146908 & -1.5 & 8,750 $\pm$ 500 & 2.72 $\pm$ 0.08 & 39.10 $\pm$ 0.49 & 16.84 & 11\\
BHB-15 & 78.550217 & -40.069916 & -1.5 & 8,250 $\pm$ 500 & 3.24 $\pm$ 0.10 & 43.78 $\pm$ 0.93 & 18.42 & 18\\
BHB-16 & 78.567359 & -40.030395 & -1.5 & 8,500 $\pm$ 500 & 3.01 $\pm$ 0.09 & 42.66 $\pm$ 0.89 & 5.76 & 16\\
BHB-17 & 78.510361 & -40.022594 & -1.5 & 7,250 $\pm$ 500 & 4.23 $\pm$ 0.15 & 44.75 $\pm$ 0.69 & 5.66 & 11\\
BHB-18 & 78.517273 & -40.079132 & -1.5 & 8,750 $\pm$ 500 & 2.81 $\pm$ 0.08 & 41.62 $\pm$ 0.45 & 7.89 & 11\\
BHB-19 & 78.541298 & -40.003937 & -1.5 & 7,500 $\pm$ 500 & 3.79 $\pm$ 0.13 & 40.96 $\pm$ 0.64 & 19.19 & 11\\
BHB-20 & 78.504898 & -40.031822 & -1.5 & 8,250 $\pm$ 500 & 3.16 $\pm$ 0.09 & 41.79 $\pm$ 0.31 & 10.48 & 11\\
BHB-21 & 78.505676 & -40.025120 & -1.0 & 9,750 $\pm$ 500 & 2.08 $\pm$ 0.06 & 35.38 $\pm$ 0.96 & 7.98 & 18\\
BHB-22 & 78.550384 & -40.020740 & -0.5 & 9,750 $\pm$ 500 & 2.13 $\pm$ 0.06 & 36.06 $\pm$ 0.99 & 12.51 & 18\\
BHB-23 & 78.488594 & -40.040214 & -1.5 & 10,250 $\pm$ 500 & 1.84 $\pm$ 0.05 & 33.38 $\pm$ 0.41 & 12.74 & 11\\
BHB-24 & 78.546638 & -40.026207 & -0.5 & 9,750 $\pm$ 500 & 2.13 $\pm$ 0.06 & 36.94 $\pm$ 0.32 & 4.03 & 11\\
BHB-25 & 78.521225 & -40.019520 & -1.5 & 9,750 $\pm$ 500 & 2.21 $\pm$ 0.06 & 40.10 $\pm$ 0.45 & 11.70 & 11\\
BHB-26 & 78.426865 & -40.154755 & -1.0 & 9,750 $\pm$ 500 & 2.09 $\pm$ 0.06 & 35.71 $\pm$ 0.52 & 11.14 & 11\\
BHB-27 & 78.517349 & -40.050198 & -1.5 & 9,250 $\pm$ 500 & 2.38 $\pm$ 0.07 & 37.23 $\pm$ 0.64 & 23.06 & 7\\
BHB-28 & 78.520325 & -40.054283 & -1.0 & 9,500 $\pm$ 500 & 2.44 $\pm$ 0.07 & 43.71 $\pm$ 0.56 & 22.87 & 7\\
BHB-29 & 78.509880 & -40.052067 & -0.5 & 9,750 $\pm$ 500 & 2.12 $\pm$ 0.05 & 36.77 $\pm$ 0.39 & 27.27 & 7\\
BHB-30 & 78.525955 & -40.042282 & -0.5 & 9,500 $\pm$ 500 & 2.33 $\pm$ 0.06 & 39.84 $\pm$ 0.41 & 21.87 & 7\\
\hline
BHB-31 & 78.520195 & -40.040668 & -1.0 & 9,500 $\pm$ 500 & 2.54 $\pm$ 0.07 & 47.57 $\pm$ 0.62 & 17.73 & 7\\
BHB-32 & 78.518440 & -40.040531 & -1.5 & 10,000 $\pm$ 500 & 2.02 $\pm$ 0.05 & 36.77 $\pm$ 0.34 & 7.933 & 7\\
BHB-33 & 78.530701 & -40.040970 & -0.5 & 9,000 $\pm$ 500 & 2.60 $\pm$ 0.07 & 40.09 $\pm$ 0.63 & 23.74 & 7\\
BHB-34 & 78.553841 & -40.048237 & -1.0 & 9,000 $\pm$ 500 & 2.76 $\pm$ 0.08 & 45.15 $\pm$ 0.63 & 24.02 & 7\\
BHB-35 & 78.515976 & -40.051266 & -1.0 & 9,750 $\pm$ 500 & 2.14 $\pm$ 0.06 & 37.36 $\pm$ 0.50 & 16.29 & 7\\
BHB-36 & 78.552643 & -40.055920 & -1.0 & 10,500 $\pm$ 500 & 1.80 $\pm$ 0.04 & 35.43 $\pm$ 0.32 & 10.53 & 7\\
BHB-37 & 78.532608 & -40.046638 & -1.5 & 8,750 $\pm$ 500 & 2.77 $\pm$ 0.08 & 40.63 $\pm$ 0.81 & 18.83 & 7\\
BHB-38 & 78.539772 & -40.063961 & -1.0 & 9,500 $\pm$ 500 & 2.30 $\pm$ 0.06 & 38.88 $\pm$ 0.45 & 11.79 & 7\\
BHB-39 & 78.518684 & -40.041420 & -0.5 & 9,250 $\pm$ 500 & 2.48 $\pm$ 0.07 & 40.76 $\pm$ 0.40 & 35.73 & 7\\
BHB-40 & 78.521194 & -40.038918 & -1.0 & 9,000 $\pm$ 500 & 2.61 $\pm$ 0.07 & 40.40 $\pm$ 0.66 & 19.24 & 7\\
BHB-41 & 78.525856 & -40.032352 & -0.5 & 9,000 $\pm$ 500 & 2.58 $\pm$ 0.07 & 39.26 $\pm$ 0.43 & 31.53 & 7\\
BHB-42 & 78.527428 & -40.028969 & -1.5 & 10,000 $\pm$ 500 & 2.04 $\pm$ 0.05 & 37.49 $\pm$ 0.36 & 10.07 & 7\\
BHB-43 & 78.522675 & -40.048580 & -1.5 & 8,750 $\pm$ 500 & 3.13 $\pm$ 0.02 & 50.97 $\pm$ 0.95 & 12.11 & 7\\
BHB-44 & 78.533165 & -40.044201 & -0.5 & 9,500 $\pm$ 500 & 2.53 $\pm$ 0.07 & 46.93 $\pm$ 0.71 & 13.98 & 7\\
BHB-45 & 78.528137 & -40.053436 & -1.0 & 10,250 $\pm$ 500 & 1.92 $\pm$ 0.05 & 36.54 $\pm$ 0.49 & 6.464 & 7\\
BHB-46 & 78.522995 & -40.049427 & -0.5 & 10,500 $\pm$ 500 & 1.84 $\pm$ 0.05 & 36.99 $\pm$ 0.76 & 3.000 & 7\\
BHB-47 & 78.533607 & -40.055218 & -1.0 & 10,250 $\pm$ 500 & 1.91 $\pm$ 0.05 & 36.17 $\pm$ 0.42 & 11.39 & 7\\
BHB-48 & 78.540390 & -40.063343 & -0.5 & 9,500 $\pm$ 500 & 2.43 $\pm$ 0.06 & 43.46 $\pm$ 0.51 & 20.80 & 7\\
BHB-49 & 78.530136 & -40.048267 & -1.0 & 8,750 $\pm$ 500 & 2.98 $\pm$ 0.09 & 47.06 $\pm$ 0.73 & 33.96 & 7\\
BHB-50 & 78.533630 & -40.045277 & -0.5 & 9,750 $\pm$ 500 & 2.41 $\pm$ 0.06 & 47.36 $\pm$ 0.55 & 30.61 & 7\\
BHB-51 & 78.530373 & -40.058037 & -0.5 & 9,000 $\pm$ 500 & 2.79 $\pm$ 0.08 & 45.99 $\pm$ 0.67 & 19.43 & 7\\
BHB-52 & 78.528709 & -40.053467 & -1.0 & 10,250 $\pm$ 500 & 1.92 $\pm$ 0.05 & 36.54 $\pm$ 0.49 & 6.464 & 7\\
BHB-53 & 78.517830 & -40.045273 & -1.5 & 10,250 $\pm$ 500 & 1.92 $\pm$ 0.05 & 36.88 $\pm$ 0.43 & 7.511 & 7\\
BHB-54 & 78.533768 & -40.043423 & -1.5 & 10,250 $\pm$ 500 & 1.89 $\pm$ 0.05 & 35.60 $\pm$ 0.34 & 9.490 & 7\\
BHB-55 & 78.535095 & -40.046638 & -0.5 & 9,750 $\pm$ 500 & 2.08 $\pm$ 0.05 & 35.48 $\pm$ 0.51 & 16.28 & 7\\
BHB-56 & 78.531456 & -40.036232 & -1.0 & 9,750 $\pm$ 500 & 2.11 $\pm$ 0.05 & 36.39 $\pm$ 0.38 & 18.68 & 7\\
BHB-57 & 78.522224 & -40.053318 & -1.0 & 9,500 $\pm$ 500 & 2.28 $\pm$ 0.06 & 38.23 $\pm$ 0.54 & 22.80 & 7\\
BHB-58 & 78.532806 & -40.048916 & -1.5 & 9,500 $\pm$ 500 & 2.17 $\pm$ 0.06 & 34.48 $\pm$ 0.61 & 23.30 & 7\\
BHB-59 & 78.504097 & -40.039898 & -0.5 & 9,000 $\pm$ 500 & 2.86 $\pm$ 0.08 & 48.60 $\pm$ 0.59 & 26.56 & 7\\
BHB-60 & 78.544830 & -40.043510 & -1.0 & 9,000 $\pm$ 500 & 2.86 $\pm$ 0.08 & 48.54 $\pm$ 0.64 & 22.22 & 7\\
BHB-61 & 78.524002 & -40.050644 & -0.5 & 9,250 $\pm$ 500 & 2.42 $\pm$ 0.07 & 38.75 $\pm$ 0.49 & 19.71 & 7\\
BHB-62 & 78.523346 & -40.049740 & -0.5 & 10,500 $\pm$ 500 & 1.84 $\pm$ 0.05 & 36.99 $\pm$ 0.76 & 3.000 & 7\\
BHB-63 & 78.534225 & -40.046970 & -0.5 & 9,500 $\pm$ 500 & 2.34 $\pm$ 0.06 & 40.16 $\pm$ 0.75 & 3.533 & 7\\
BHB-64 & 78.544800 & -40.043964 & -1.0 & 9,000 $\pm$ 500 & 2.86 $\pm$ 0.08 & 48.54 $\pm$ 0.64 & 22.22 & 7\\
BHB-65 & 78.530678 & -40.040634 & -0.5 & 9,000 $\pm$ 500 & 2.60 $\pm$ 0.07 & 40.09 $\pm$ 0.62 & 23.74 & 7\\
BHB-66 & 78.541359 & -40.027889 & -1.5 & 10,000 $\pm$ 500 & 1.98 $\pm$ 0.05 & 35.54 $\pm$ 0.33 & 9.603 & 7\\
BHB-67 & 78.520767 & -40.044174 & -1.5 & 10,250 $\pm$ 500 & 1.94 $\pm$ 0.05 & 37.55 $\pm$ 0.39 & 18.55 & 7\\
BHB-68 & 78.525414 & -40.062984 & -1.0 & 9,500 $\pm$ 500 & 2.39 $\pm$ 0.06 & 42.05 $\pm$ 0.49 & 20.63 & 7\\
BHB-69 & 78.541512 & -40.051468 & -1.5 & 10,000 $\pm$ 500 & 2.03 $\pm$ 0.05 & 37.33 $\pm$ 0.37 & 9.168 & 7\\
BHB-70 & 78.553497 & -40.051537 & -0.5 & 10,250 $\pm$ 500 & 1.86 $\pm$ 0.05 & 34.61 $\pm$ 0.34 & 22.37 & 7\\
\hline
BHB-71 & 78.543266 & -40.047127 & -0.5 & 9,000 $\pm$ 500 & 2.60 $\pm$ 0.07 & 40.05 $\pm$ 0.65 & 24.29 & 7\\
\hline
BHBd-1 $^{4}$ & 78.494659 & -40.038635 & 0.5 & 12,250 $\pm$ 500 & 1.22 $\pm$ 0.03 & 30.38 $\pm$ 0.51 & 28.80 & 11\\
BHBd-2 $^{4}$ & 78.482582 & -40.042618 & 0.0 & 11,750 $\pm$ 500 & 1.37 $\pm$ 0.03 & 32.19 $\pm$ 0.92 & 28.07 & 16\\
BHBd-3 $^{4}$ & 78.548469 & -40.019894 & 0.2 & 11,750 $\pm$ 500 & 1.36 $\pm$ 0.03 & 31.98 $\pm$ 0.90 & 22.70 & 15\\
BHBd-4 & 78.498787 & -40.089081 & 0.5 & 11,000 $\pm$ 500 & 1.55 $\pm$ 0.04 & 31.63 $\pm$ 0.85 & 14.67 & 16\\
BHBd-5 & 78.509033 & -40.065044 & -0.5 & 11,250 $\pm$ 500 & 1.50 $\pm$ 0.03 & 32.61 $\pm$ 0.37 & 19.34 & 11\\
BHBd-6 & 78.476257 & -40.065121 & 0.2 & 11,000 $\pm$ 500 & 1.65 $\pm$ 0.04 & 36.01 $\pm$ 0.94 & 17.62 & 16\\
BHBd-7 & 78.495964 & -40.0658493 & 0.2 & 11,000 $\pm$ 500 & 1.56 $\pm$ 0.04 & 32.02 $\pm$ 0.84 & 23.76 & 18\\
BHBd-8 & 78.545028 & -40.092133 & 0.5 & 11,500 $\pm$ 500 & 1.61 $\pm$ 0.04 & 40.72 $\pm$ 0.38 & 29.77 & 11\\
BHBd-9 & 78.554802 & -40.064109 & -0.5 & 11,000 $\pm$ 500 & 1.63 $\pm$ 0.04 & 35.04 $\pm$ 0.98 & 35.65 & 16\\
BHBd-10 & 78.572716 & -40.017147 & 0.0 & 11,750 $\pm$ 500 & 1.38 $\pm$ 0.03 & 32.61 $\pm$ 0.56 & 15.06 & 11\\
BHBd-11 & 78.532326 & -40.040825 & 0.0 & 11,750 $\pm$ 500 & 1.29 $\pm$ 0.03 & 28.85 $\pm$ 0.28 & 9.18 & 7\\
BHBd-12 & 78.531876 & -40.046726 & 0.2 & 11,250 $\pm$ 500 & 1.45 $\pm$ 0.03 & 30.60 $\pm$ 0.36 & 29.04 & 7\\
BHBd-13 & 78.533455 & -40.047199 & 0.0 & 11,500 $\pm$ 500 & 1.49 $\pm$ 0.04 & 34.87 $\pm$ 0.81 & 2.011 & 7\\
BHBd-14 & 78.528633 & -40.041973 & -0.5 & 11,000 $\pm$ 500 & 1.56 $\pm$ 0.04 & 32.38 $\pm$ 0.32 & 8.56 & 7\\
BHBd-15 & 78.528740 & -40.037434 & 0.0 & 11,500 $\pm$ 500 & 1.43 $\pm$ 0.03 & 32.07 $\pm$ 0.33 & 15.91 & 7\\
BHBd-16 & 78.526413 & -40.047501 & 0.0 & 11,750 $\pm$ 500 & 1.31 $\pm$ 0.03 & 29.37 $\pm$ 0.45 & 7.828 & 7\\
BHBd-17 & 78.530174 & -40.050919 & 0.2 & 11,250 $\pm$ 500 & 1.44 $\pm$ 0.03 & 29.99 $\pm$ 0.36 & 14.45 & 7\\
BHBd-18 & 78.526314 & -40.058327 & -0.5 & 11,000 $\pm$ 500 & 1.71 $\pm$ 0.04 & 38.72 $\pm$ 0.40 & 12.04 & 7\\
BHBd-19 & 78.538643 & -40.047455 & 0.0 & 11,500 $\pm$ 500 & 1.43 $\pm$ 0.03 & 32.11 $\pm$ 0.34 & 8.924 & 7\\
BHBd-20 & 78.528465 & -40.041519 & -0.5 & 11,000 $\pm$ 500 & 1.57 $\pm$ 0.04 & 32.38 $\pm$ 0.32 & 8.562 & 7\\
BHBd-21 & 78.528572 & -40.049442 & 0.0 & 11,000 $\pm$ 500 & 1.55 $\pm$ 0.04 & 31.62 $\pm$ 0.25 & 16.02 & 7\\
BHBd-22 & 78.520851 & -40.047207 & -0.5 & 11,000 $\pm$ 500 & 1.57 $\pm$ 0.04 & 32.36 $\pm$ 0.28 & 13.01 & 7\\
\hline
\hline 
\end{longtable*}

We have shown the SEDs of only those BHB and BHBd stars for which spectroscopic determination of temperatures are available \citep{2012A&A...539A..19G}, so as to compare the temperature estimated from both the procedures and validate the estimations from the SED fits. 

\begin{figure*}[ht!]
\epsscale{1.25}
\plotone{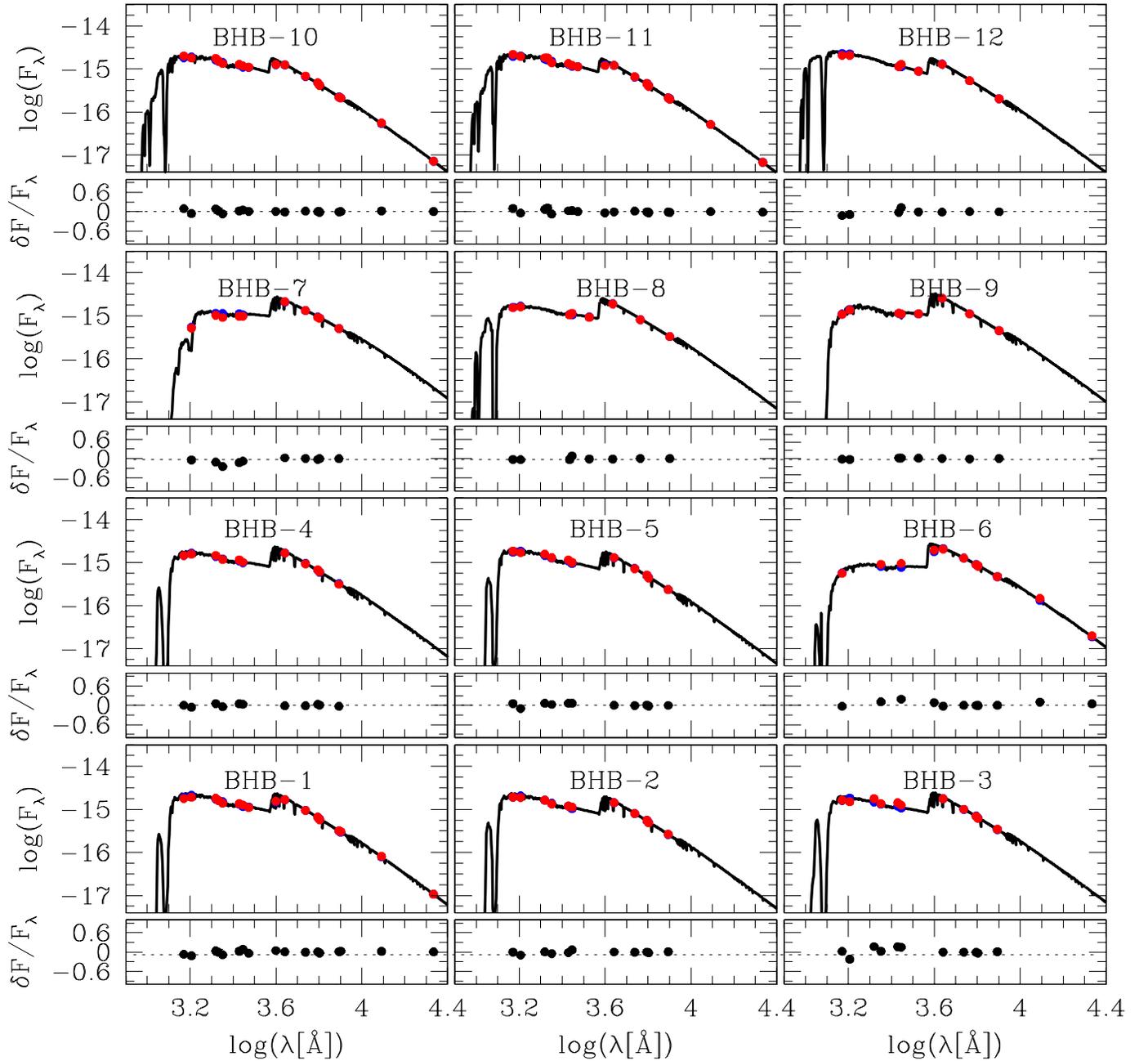}
\caption{The SED of the BHB stars are plotted in the $upper$ $panel$ and the corresponding residuals are plotted in the $bottom$ $panel$. Blue dots denote the synthetic flux and red dots denote the observed flux. The SED fits show a good match of the theoretical and observed fluxes in all the wavelength bands. \label{fig:nbhb}}
\end{figure*}

\begin{figure*}[ht!]
\epsscale{1.25}
\plotone{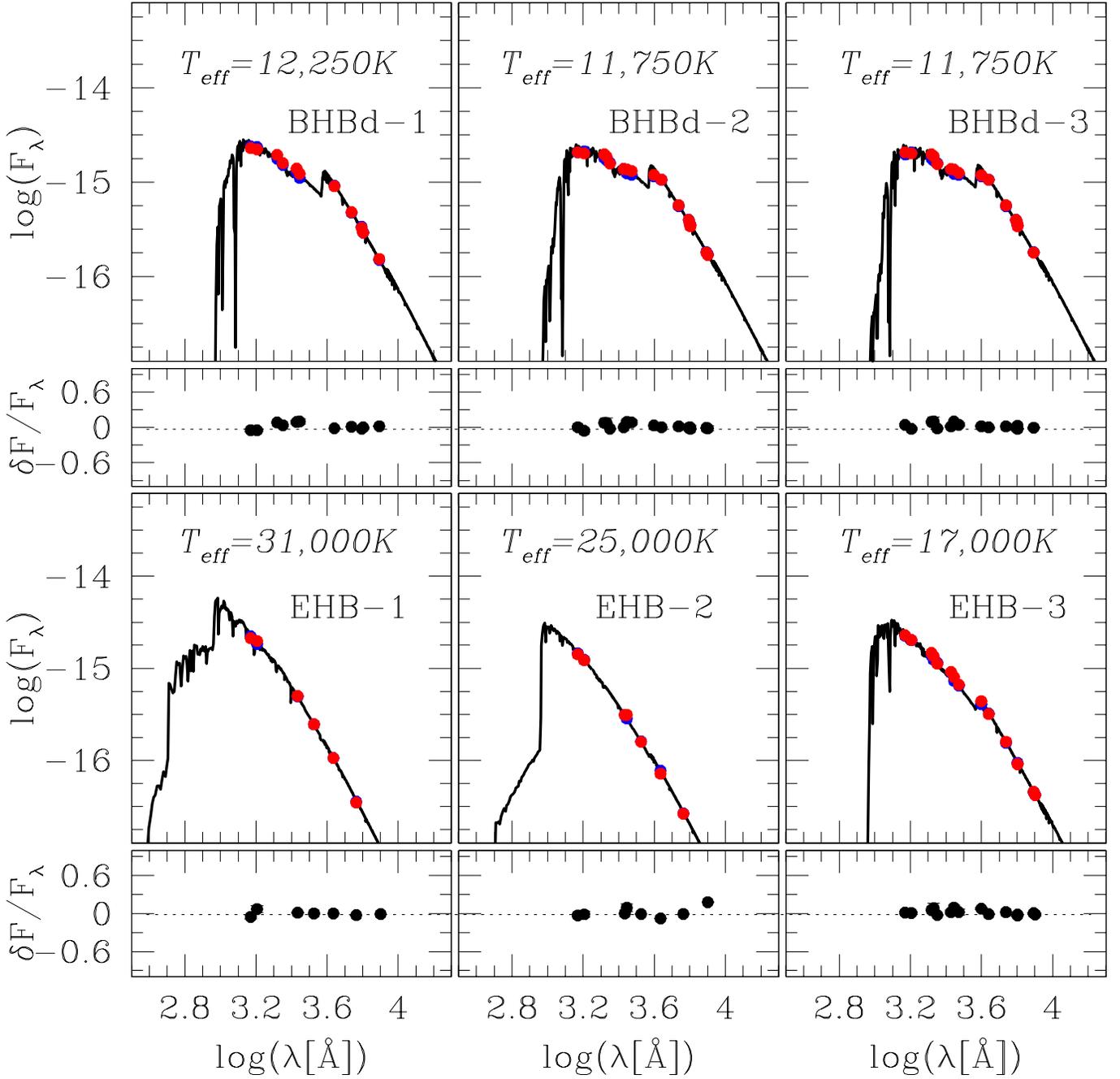}
\caption{The SED of the BHBd stars are plotted in the $upper$ $panel$ and the corresponding residuals are plotted in the $bottom$ $panel$. The SED fit of the EHB-1, EHB-2 and EHB-3 stars in logarithmic scale are shown in the $lower$ $panel$. Blue dots denote the synthetic flux and red dots denote the observed flux. The estimated temperatures are also indicated in the figures.\label{fig:hook}}
\end{figure*}

\begin{figure*}[ht!]
\epsscale{1.25}
\plotone{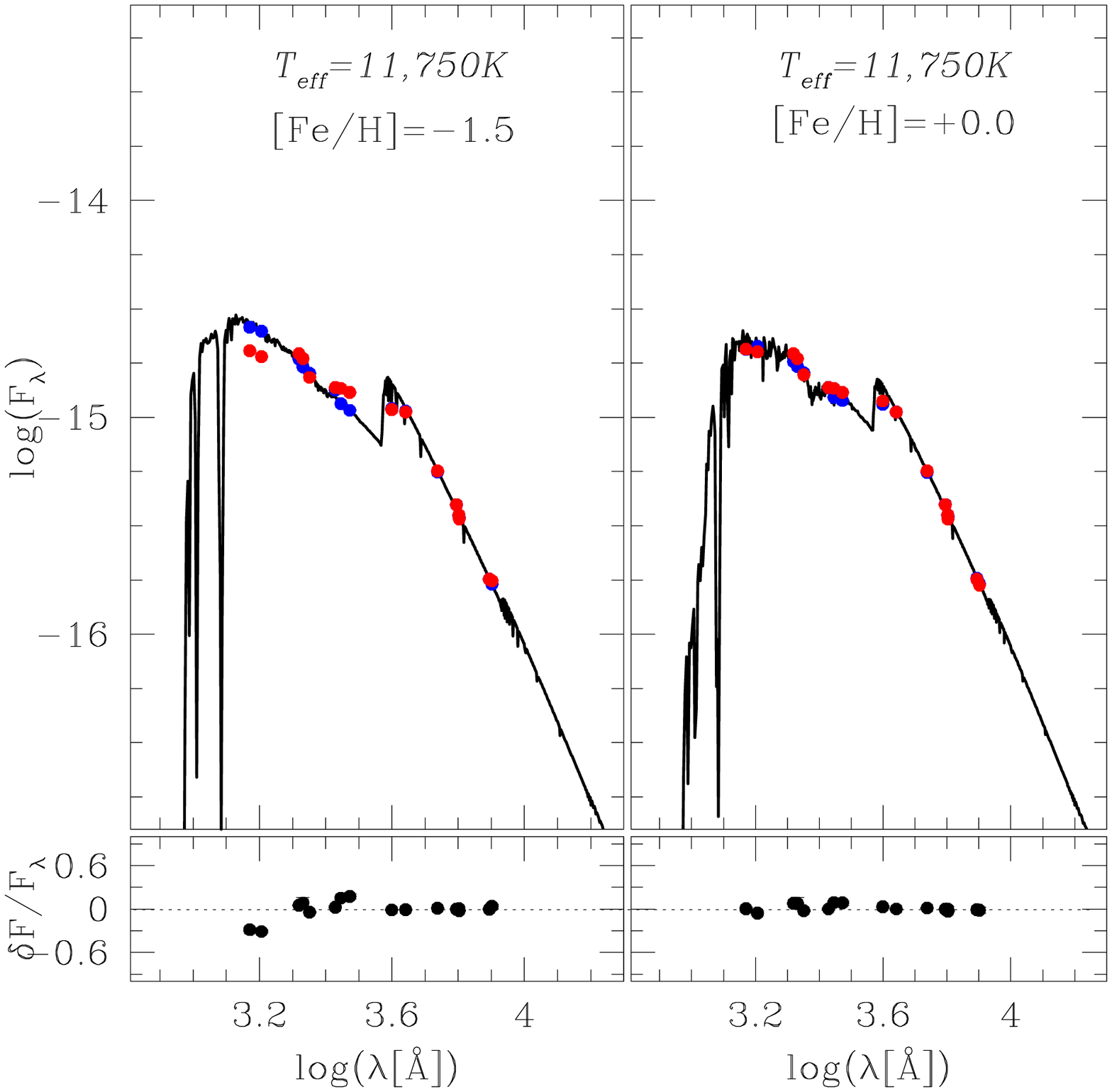}
\caption{ The SED of BHBd-2 star when fitted with the cluster metallicity show a decrease in FUV flux, as shown in the $left$ $panel$. The SED shows a better fit at a higher metallicity, as shown in the $right$ $panel$.}\label{fig:diffusion}
\end{figure*}

In Figure \ref{fig:nbhb}, we have shown the SEDs and the corresponding residual plot for all the BHB stars having spectroscopically determined temperatures. The observed flux in the SEDs are fitted with the Castelli/Kurucz stellar atmospheric models. The residual plot, shown below each of the SED, shows the difference between the observed flux and the synthetic flux normalised with respect to the observed flux, corresponding to the flux measurements in each pass-band. The residual plots show a good match of the theoretical and observed fluxes in all the wavelength bands, with the residual values close to zero. The SEDs for the rest of the BHB stars are also fitted with models to estimate the parameters, though the plots are not shown.

In the $upper$ $panels$ of Figure \ref{fig:hook}, we show the SEDs and the corresponding residual plots of BHBd stars for which the spectroscopic estimation of temperatures are available. The SEDs of these stars are fitted with higher metallicity models. In Figure \ref{fig:diffusion}, the SED of BHBd-2 star is shown as an example. The SED fit shows a significant drop in FUV flux, when fitted with atmospheric models of cluster metallicity. The SED is found to be better fitted with a model of higher metallicity. This behaviour has been observed for all the BHBd stars found in our sample.

We fit the SEDs of three EHB stars located in the hot end of the UV CMDs, as shown in the $lower$ $panel$ of the Figure \ref{fig:hook}. The SEDs of EHB-1 and EHB-2 stars located in the inner region are fitted with models to obtain ($T_{eff}$ = 31,000 $\pm$ 500 K, $R/R_{\odot}$ = 0.16 $\pm$ 0.01, $L/L_{\odot}$ = 20.50 $\pm$ 0.24 ) and ($T_{eff}$ = 25,000 $\pm$ 500 K, $R/R_{\odot}$ = 0.17 $\pm$ 0.02, $L/L_{\odot}$ = 11.53 $\pm$ 0.13) respectively.

The residual plot of EHB-2 star shows that the SED is not fitted well towards the longer wavelength, with the presence of excess flux in the F814W band. As this is HST flux with very small errors, the flux deviation is statistically significant. The SED of EHB-3 star is fitted with a temperature of $\sim$ 17,000 K ($T_{eff}$ = 17,000 $\pm$ 500 K, $R/R_{\odot}$ = 0.52 $\pm$ 0.03, $L/L_{\odot}$ = 20.33 $\pm$ 0.66). The residual plots show SEDs of EHB-1 and EHB-3 are fitted with higher metallicity models, similar to that of the BHBd stars.

\begin{figure*}[ht!]
\epsscale{1.05}
\plotone{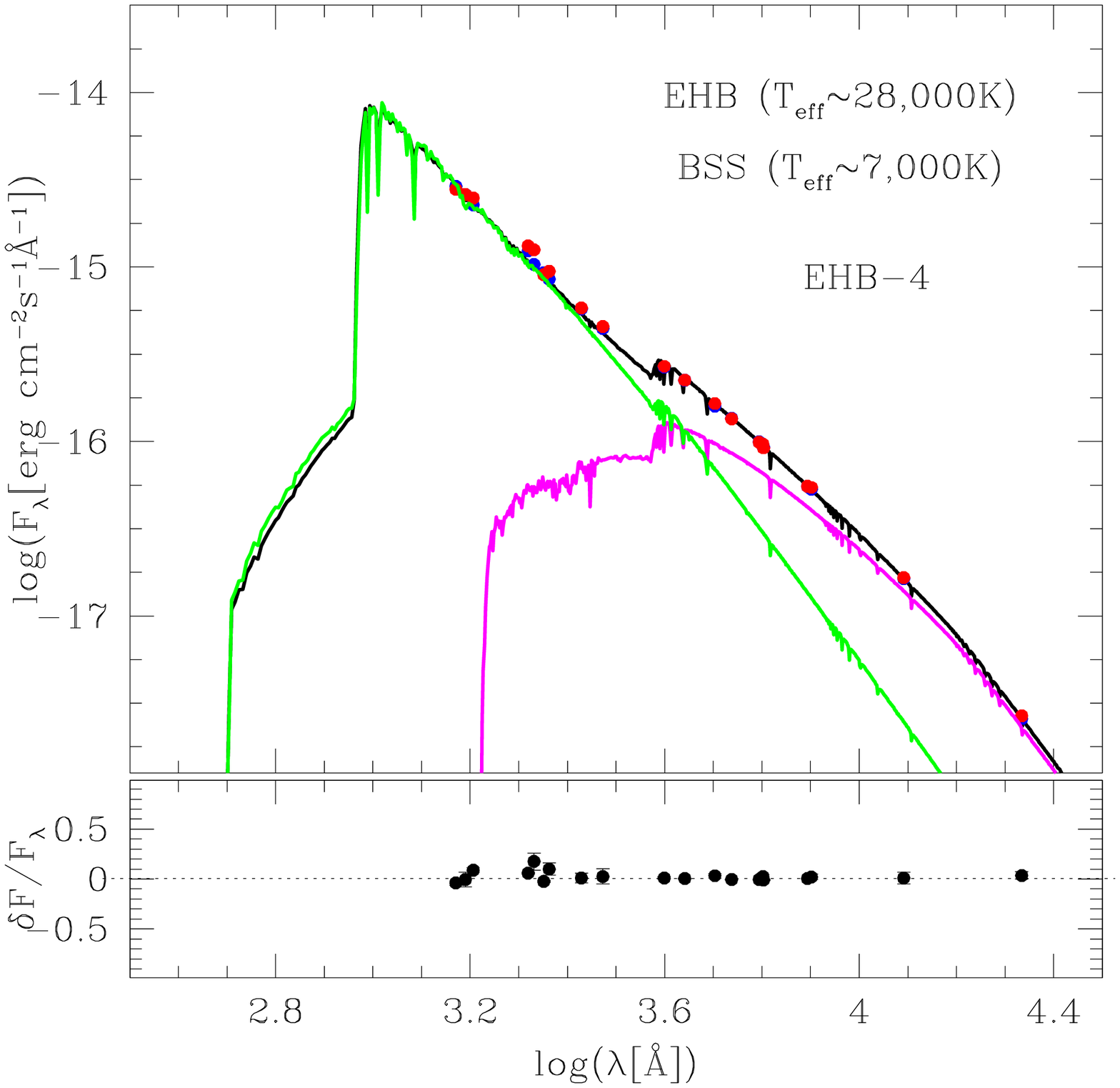}
\caption{The SED of the EHB-4 star is plotted. The SED of this star cannot be fitted with a single SED and the best fit results in a composite SED fit of stars with a BSS ($T_{eff}$ $\sim$ 7,000 K) and EHB ($T_{eff}$ $\sim$ 28,000 K), and are shown with magenta and green colors, respectively.\label{fig:comp}}
\end{figure*}

The SED of EHB-4 star as shown in Figure \ref{fig:comp}, located in the outer region, cannot be fitted with a single SED. A hotter spectrum leaves a significant IR excess and a cooler spectrum leaves a significant UV excess. 
In order to obtain the best fit for the entire SED, we fit the SED with a composite spectra using Kurucz stellar atmospheric models. The best fit ($\chi^{2}_{reduced}$ = 6.33) is obtained for a composite fit of SEDs ($T_{eff}$ = 7,000 $\pm$ 500 K, $R/R_{\odot}$ = 1.17 $\pm$ 0.01 and $T_{eff}$ = 28,000 K$\pm$ 500 K, $R/R_{\odot}$ = 0.21 $\pm$ 0.01). The residual plot shows that the residual flux are within 3-$\sigma$ error of the data.

The fitting parameters for the EHB stars are listed in Table 3. 

\begin{table*}
\renewcommand{\thetable}{\arabic{table}}
\centering
\caption{SED fit parameters of the EHB stars with a varying metallicity. Columns 5, 6 and 7 present the values of temperatures, radii and luminosities of EHBs derived from the SED fitting.} \label{tab:decimal}
\begin{tabular}{ccccccccc}
\tablewidth{0pt}
\hline
\hline
Star ID & RA (J2000) & DEC (J2000) & [Fe/H] & $T_{eff}$ & $R$/$R_{\sun}$ & $L$/$L_{\sun}$ & $\chi^{2}_{reduced}$ & $N_{fit}$\\
& [$\degr$] & [$\degr$] & & [K] & & & &\\\hline
%%\decimals
EHB-1 & 78.518341 & -40.045078 & 0.5 & 31,000 $\pm$ 500 & 0.16 $\pm$ 0.01 & 20.50 $\pm$ 0.24 & 10.78 & 7\\
EHB-2 & 78.507988 & -40.042366 & -1.5 & 25,000 $\pm$ 500 & 0.17 $\pm$ 0.02 & 11.53 $\pm$ 0.13 & 479.40 & 8 \\
EHB-3 & 78.567474 & -40.032307 & 0.0 & 17,000 $\pm$ 500 & 0.52 $\pm$ 0.03 & 20.33 $\pm$ 0.66 & 18.94 & 15\\
\hline
EHB-4 & 78.572006 & -39.993195 & -0.5 & 28,000 $\pm$ 500 & 0.21 $\pm$ 0.01 & 24.44 $\pm$ 0.93 & 6.33 & 20 \\
& & & -1.5 & 7,000 $\pm$ 500 & 1.17 $\pm$ 0.01 & 2.95 $\pm$ 0.20 & &\\
\hline
\hline 
\end{tabular}
\end{table*}

\begin{figure}[ht!]
\epsscale{1.25}
\plotone{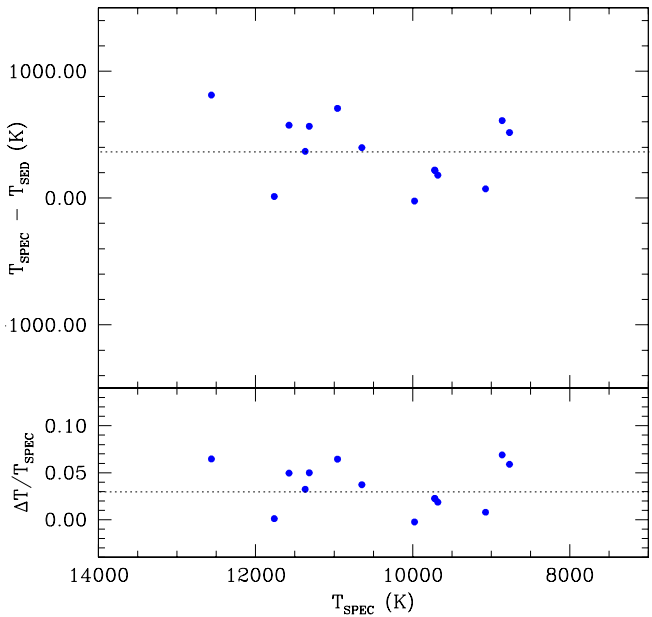}
\caption{The difference of spectroscopic and the temperatures derived through SED are shown. The residuals indicate that the SED derived temperatures are estimated within error of $\Delta$ $T_{eff}$ $\sim$ 500 k.\label{fig:diff}}
\end{figure}

We also compared the difference in temperatures estimated from SED fit to the spectroscopically determined temperatures. Figure \ref{fig:diff} shows the difference of the temperatures determined through SED and spectroscopy. The difference in temperature was found to have an offset of $\Delta T$ $\sim$ 373 $\pm$ 266 K, and $\Delta T$/$T_{SPEC}$ $\sim$ 0.03 $\pm$ 0.02 is the fractional residual of the difference. The residual plots show that the temperatures determined through SED match the temperatures determined using spectroscopic method within error limits.

\subsection{Radius vs temperature and temperature vs luminosity diagrams} \label{subsec:rad}
The parameters, $R$/$R_{\odot}$ and $L$/$L_{\odot}$ estimated from SEDs are compared with the Flexible Stellar Population Synthesis (FSPS) models of \cite{2009ApJ...699..486C}, generated for a metallicity value of [Fe/H] = $-$1.2 dex \citep{2013AJ....145...25K}, with a distance modulus of 15.52 mag and an assumed cluster age of 10 Gyr \citep{2008ApJ...672L.115C}. We constructed the radius-temperature and temperature-luminosity plots to understand and compare the parameters for the HB sequence. 

\begin{figure*}[ht!]
\epsscale{1.05}
\plotone{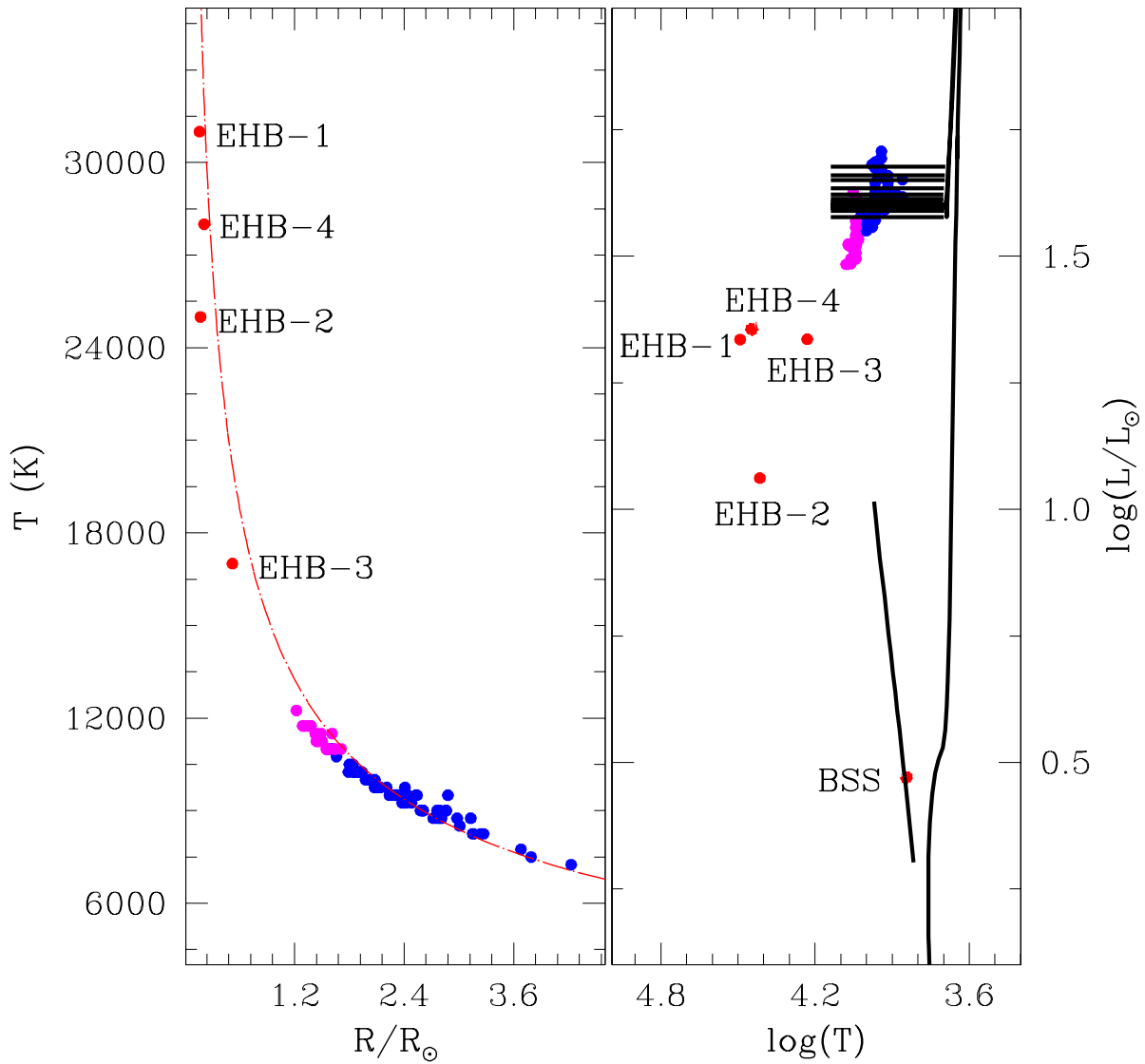}
\caption{The $left$ $panel$ shows radius-temperature relation of the stars for which SED fit parameters have been estimated. The plot shows a good match for the BHB stars. The $right$ $panel$ shows the temperature-luminosity relationship of all the SED fit stars. The plot shows deviation starting from the canonical HB track for the BHBd stars. The EHB star also shows deviation from the canonical track, and this deviation is much more pronounced for the EHB-2 and EHB-3 stars. The fit parameters of the companion to EHB-4 star suggest it to be a BSS. \label{fig:par}}
\end{figure*}

In the $left$ $panel$ of Figure \ref{fig:par}, the radius-temperature relationship is plotted. By considering the mean value of luminosity and radius of the BHB stars ($L$/$L_{\sun}$ $\sim$ 38.63 and $R$/$R_{\sun}$ $\sim$ 2.25), we have plotted the constant luminosity curve with a temperature ranging from 4,000 K to 40,000 K using the relationship; $L$ $=$ $4\pi r^{2} \sigma T^{4}$, where $\sigma$ is the Stefan-Boltzmann constant and $T$ is the surface temperature of the star. The BHB stars fitted quite well in the radius-temperature plot. The plot, however, shows deviation from the constant luminosity curve for the BHBd stars and the EHB-2 star. We also see a significant deviation for EHB-3 star. We note that the EHB-3 star is located in between the BHBd stars and the other EHB stars, with respect to T$_{eff}$ and radius. The radius of the EHB-3 is about half the radius of the BHB stars, and about 2.5 times the radius of the other EHB stars. The BHBd stars showing deviation from the constant luminosity curve are expected to have a low-surface gravity, due to radiative levitation of heavy elements \citep{1999ApJ...524..242G,1995A&A...294...65M}. \cite{1995A&A...294...65M} found that log$g$ for these stars could be underestimated by at most 0.1 dex, while \cite{1997A&A...320..257L} found that for helium-weak stars, log$g$ could be underestimated by a value of $\sim$ 0.25 dex if a solar metallicity model were used.

\begin{figure*}[ht!]
\epsscale{1.05}
\plotone{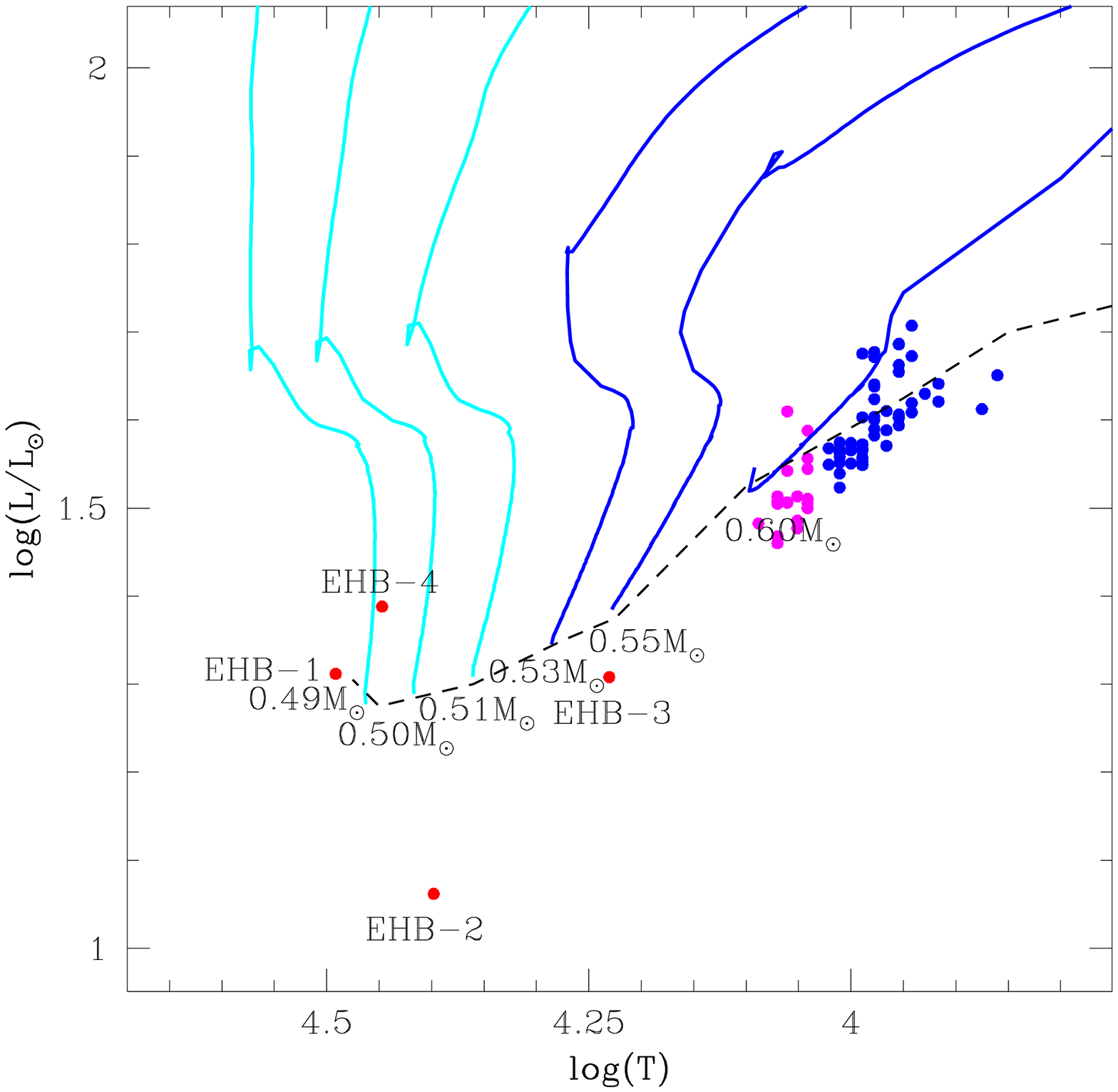}
\caption{The SED fit parameters are plotted along with the stellar evolutionary sequences of the Post-ZAHB. Black dashed line indicates the location of
the ZAHB and is shown with an interval of 5 Myr, while each step of the post-HB evolutionary track is plotted with a step of 0.5 Myr. The blue and Cyan symbols indicate the sequences that populate the HB sequence's blue and hot end. \label{fig:lum}}
\end{figure*}

The temperature-luminosity plot in the $right$ $panel$ of Figure \ref{fig:par} shows that the BHB stars have an almost constant bolometric luminosity. The plot diverges from the canonical evolutionary model for the BHBd stars. The EHB stars are significantly fainter, with similar luminosity for EHB-1, EHB-2 and the hot component of EHB-4, whereas EHB-3 is found to be significantly fainter. The companion to EHB-4 star is found to be located in the BSS sequence confirming it to be a BSS, which has a radius of $R$/$R_{\sun}$ = 1.17$\pm$0.01 and $L$/$L_{\sun}$ = 2.95 $\pm$0.20. 

To compare the SED fit parameters with the core mass of the EHB population, we plot the H-R diagram using the parameters estimated from the SED, along with the stellar evolutionary track of ZAHB and Post-ZAHB stars using BASTI-IAC model \citep{Hidalgo_2018} as shown in Figure \ref{fig:lum}.  The location of EHB-1 star is on the left side of the lowest theoretical post-ZAHB evolutionary model with core mass of 0.4920 $M_{\odot}$. This suggests that EHB-1 star will evolve to a white-dwarf (WD) stage from ZAHB track. The HB stars having very low envelope masses go directly to white dwarf phase, while the HB stars having envelope masses larger than 0.02 $M_{\odot}$ can partly evolve to the AGB phase \citep{2019A&A...627A..34M,1990ApJ...364...35G}. The other two EHB stars likely follow the blue and cyan curve, respectively. The EHB-2 star lies below the ZAHB track of the core mass of $\sim$ 0.50 $M_{\odot}$, but is significantly less luminous. The EHB-3 falls on the ZAHB curve with a core mass of $\sim$ 0.55 $M_{\odot}$. The EHB-4 is located at the Post-ZAHB track with a core mass of $\sim$ 0.49 $M_{\odot}$, and might be on the verge of post-HB evolution.

\section{Discussions} \label{subsec:diss} 
We have studied the UV bright members in NGC 1851 using the UVIT/ASTROSAT, Gaia and UVOT data. The HB members in the cluster are identified up to a large radius through optical and UV-optical CMDs. We detected 2 EHB stars each in the inner and outer regions. These stars, along with the BHBd stars are characterised using the SEDs. Here we discuss the implications of these findings.

\begin{enumerate} 
\item The SEDs of BHBd stars show a decrease in FUV flux when fitted with atmospheric models of cluster metallicity. These SEDs are better fitted with higher metallicity models, which satisfactorily reproduce the observed UV fluxes, as expected from atomic diffusion \citep{1999ApJ...524..242G,2016ApJ...822...44B,2019A&A...627A..34M}. \cite{moehler2000hot} and \cite{2019A&A...627A..34M} used solar metallicity LTE model spectra to incorporate the effects of radiative levitation. \cite{1999ApJ...524..242G}  obtained the photometric evidence of atomic diffusion for the BHB stars hotter than $\sim$ 11,500 K. They found that these stars are somewhat fainter ($\sim$ 0.1 mag) in UIT/1620 $\AA$ and WFPC2/F160BW, but are typically brighter in Str\"{o}mgren $\textit{u}$ than predicted from canonical HB model. They explained that atomic diffusion of heavy elements in the HB stars lying in the temperature range of 11,500 K to 20,000 K may be responsible for such variation. So, atomic diffusion may enhance the abundance of metals in the stellar atmosphere of these stars. They also identified that metal rich models could provide a better fit for the temperatures hotter than G-jump in the ultraviolet CMD. \cite{2000A&A...363..159B} and \cite{2001ApJ...562..368B} observed a similar effect in the BHB stars of NGC 2808. \cite{2019MNRAS.482.1080S} also found that HB stars, hotter than $\sim$ 11,500 K in NGC 288 showed reduced FUV flux. For the stars located between G-jump and M-jump, \cite{2016ApJ...822...44B} obtained a crude approximation of [Fe/H]=0.5 by fitting the observed CCP with the synthetic spectra with an enhanced metallicity. The results presented here are therefore consistent with the idea that there is indeed radiative levitation beyond $\sim$ 11,500 K.

\item We detect two EHB members in the inner crowded region and two in the outer region using the CMDs. Their properties estimated using the SED fits reveal interesting characteristics for all the 4 EHB stars, which are discussed below.

The EHB-1 star is fitted with $T_{eff}$ $\sim$ 31,000 K, and shows a decrease in FUV fluxes in the SED fit. This star, located at the extreme end of the HB sequence in the FUV CMD, is quite hot and could belong to the ``blue-Hook'' HB population. \cite{2000ApJ...530..352D} observed the ``blue-Hook'' in the ($\textit{160}$, $\textit{(160-V)}$) plane using UIT magnitudes, and found that these stars form a hook-like feature at the extreme end of the HB sequence in the FUV-Optical CMD. EHB-1 star has a core mass $\textless$ 0.49 $M_{\odot}$ as it is evident from the Figure \ref{fig:lum}, which also suggest that it could be a ``late-helium flasher'' or a ``blue-Hook'' star. This star is likely to evolve directly to a WD, due to it's low mass. 

The EHB-2 star is fitted with $T_{eff}$ $\sim$ 25,000 K, however, shows no variation in the FUV flux, but shows the presence of an excess flux towards the longer wavelength. It is also under luminous as seen from the H-R diagram (see right panel of Figure 14). Also, since the lower part of the spectrum is not fitted well, and it could be possible that this EHB has a cooler binary companion. This star falls below the ZAHB curve with a core mass of $\sim$ 0.50 $M_{\odot}$, which suggests that it could be a sub-luminous EHB star. There is a possibility that this EHB might have a binary origin.

The SED of EHB-3 is fitted with a temperature $\sim$ 17,000 K. This star shows a significant decrease in FUV flux in the SED, fitted over a fixed metallicity close to cluster metallicity. The SED fit reveal a better fit at higher metallicity value, as expected  The bolometric luminosity of this star show significant deviation from the canonical HB models. Also, the radius of the EHB-3 is about half the radius of the BHB stars, and about 2.5 times the radius of the other EHB stars, which suggest this star may lie at the transition region between BHB and EHB, very close to the M-jump. It has a core mass of $\sim$ 0.55 $M_{\odot}$. This star is therefore neither a true EHB nor a BHB. As this star is located in the outer region, it will be an interesting target for a spectroscopic study. 

\item The observed multi-wavelength data of EHB-4 can not be fitted with a single SED. Therefore, it is fitted with a composite SED of a hotter and a cooler spectra of temperatures $\sim$ 28000 K and 7000 K, respectively. 

The SED fit therefore indicates that this object may be a binary system. The EHB-4 is located on the outskirts of the cluster, which is a relatively less dense environment and binaries can survive in this region. The EHB component in this candidate binary system has properties similar to EHB-1, a normal EHB star. The cooler companion is found to occupy the BSS region in the H-R diagram. The finding of a BSS in the outskirts of the cluster implies that it may have formed via mass transfer mechanism. 

Binaries have been considered as a possible explanation for shaping the HB morphology of the GCs discussed by several authors \citep{2012A&A...540A..16M}. There is a well established link between field counterpart of EHB stars, i.e., sdBs (hot subluminous stars of spectral type B) and binary systems. Nearly all the sdB stars are found in close binary systems \citep{Saffer_1998,2004Ap&SS.291..321N,2004Ap&SS.291..315E,2001MNRAS.326.1391M,2020NatAs.tmp..116M}. However, the occurrence of binary EHB  stars in GCs are rare as suggested from the radial velocity studies \citep{refId0,2009A&A...498..737M}. Recently, \cite{2018OAst...27...91M} have found a main-sequence companion to an EHB star in a close binary system of the GC, NGC 6752.  

Keeping in mind the rarity of the binary systems among the EHBs in GCs, this discovery of a BSS companion to an EHB star is therefore important. Also, the identification of a BSS companion to an EHB not only provides a wonderful opportunity to search for possible EHB binary companion using UVIT/Astrosat data, but also help to understand a possible formation mechanism of EHBs as well as BSSs in GCs. If mass transfer can drive the donor to end up as an EHB star, extreme mass loss may also be able to help in the formation of EHB stars, the formation pathway of which is still debatable. However, further spectroscopic studies are needed to confirm and to estimate the detailed orbital properties and chemical signatures of this system.

\end{enumerate}
\section{SUMMARY AND CONCLUSIONS} \label{sec:summary} 

\begin{enumerate} 

\item We present a UVIT/AstroSat, UVOT/Swift, HST, Gaia-DR2 and ground based multi-wavelength study of NGC 1851. We identify the member HB stars in the cluster, covering a large radius. 

\item The optical and UV-optical CMDs, along the HST based color-color diagram is used to classify the hot HB stars in to BHB, BHBd and the EHB stars. 

\item We found HB stars hotter than the G-jump to show a decrease in the observed FUV flux with respect to the theoretical FUV fluxes when fitted with atmospheric models of cluster metallicity, however a better fit is found with higher metallicity models, as expected from the diffusion that occurs in the atmospheres of these stars.

at the metallicity close to cluster metallicity. These stars show a better SED fit for higher metallicity, as expected from the diffusion that occurs in the atmospheres of these stars.

\item We detect 4 EHB stars in this cluster with certain peculiarities. Parameters estimated using the SEDs show that the EHB-1 could be a blue-Hook star evolving to a WD. The EHB-2 is found to be a low luminosity star with the presence of IR excess in the SED. The radius of EHB-3 is found between the BHB and normal EHB stars, hence it may not be a true EHB star and may lie close to the M-jump. The SED of EHB-3 is fitted with higher metallicity model, as expected from atomic diffusion.

\item The most peculiar of the EHBs, EHB-4 ($T_{eff}$ $\sim$ 28,000 K) could be associated with the BSS as a photometric companion ($T_{eff}$ $\sim$ 7,000 K), which is an important target for follow up spectroscopic study. The less dense environment in the outer part of the GCs can be explained in terms of the post-mass transfer systems such as this candidate. This candidate EHB+BSS binary system, could help to explain the mass loss in the RGB phase, leading to the formation of EHB stars. 

\end{enumerate}

\section*{ACKNOWLEDGEMENT}
We are thankful to the reviewer for the thoughtful comments and suggestions which improved the quality of the manuscript. This work has been done during a collaborative visit to IIA, Bangalore. We are thankful to Michael H. Siegel for the useful suggestion and help during the UVOT photometric calibration. We acknowledge Hitesh lala, Yazan Momany and Simone Zaggia for critical reading of the manuscript. This publication uses the data from the AstroSat mission of the Indian Space Research Organisation (ISRO), archived at the Indian Space Science Data Centre (ISSDC). UVIT project is a result of the collaboration between IIA, Bengaluru, IUCAA, Pune, TIFR, Mumbai, several centers of ISRO, and CSA. This publication uses the data from the \textit{ASTROSAT} mission of the Indian Space Research  Organisation  (ISRO),  archived  at  the  Indian  Space  Science  Data Centre (ISSDC). This publication makes use of VOSA, developed under the Spanish Virtual Observatory project supported by the Spanish MINECO through grant AyA2017-84089. VOSA has been partially updated by using funding from the European Union's Horizon 2020 Research and Innovation Programme, under Grant Agreement $n^{0}$ 776403 (EXOPLANETS-A). This work has made use of data from the European Space Agency (ESA) mission {\it Gaia} (\url{https://www.cosmos.esa.int/gaia}), processed by the {\it Gaia} Data Processing and Analysis Consortium (DPAC, \url{https://www.cosmos.esa.int/web/gaia/dpac/consortium}). National institutions have provided funding for the DPAC, in particular, the institutions participating in the {\it Gaia} Multilateral Agreement. This research has used data, software, and/or web tools obtained from the High Energy Astrophysics Science Archive Research Center (HEASARC), a service of the Astrophysics Science Division at NASA/GSFC and of the Smithsonian Astrophysical Observatory's High Energy Astrophysics Division.
\bibliography{gaurav}

\begin{thebibliography}{}
\expandafter\ifx\csname natexlab\endcsname\relax\def\natexlab#1{#1}\fi
\providecommand{\url}[1]{\href{#1}{#1}}

\bibitem[{{Bayo} {et~al.}(2008){Bayo}, {Rodrigo}, {Barrado Y Navascu{\'e}s},
  {Solano}, {Guti{\'e}rrez}, {Morales-Calder{\'o}n}, \&
  {Allard}}]{2008A&A...492..277B}
{Bayo}, A., {Rodrigo}, C., {Barrado Y Navascu{\'e}s}, D., {et~al.} 2008, \aap,
  492, 277

\bibitem[{{Bedin} {et~al.}(2000){Bedin}, {Piotto}, {Zoccali}, {Stetson},
  {Saviane}, {Cassisi}, \& {Bono}}]{2000A&A...363..159B}
{Bedin}, L.~R., {Piotto}, G., {Zoccali}, M., {et~al.} 2000, \aap, 363, 159

\bibitem[{{Behr}(2003)}]{2003ApJS..149...67B}
{Behr}, B.~B. 2003, \apjs, 149, 67

\bibitem[{{Bianchi} {et~al.}(2011){Bianchi}, {Herald}, {Efremova}, {Girardi},
  {Zabot}, {Marigo}, {Conti}, \& {Shiao}}]{2011Ap&SS.335..161B}
{Bianchi}, L., {Herald}, J., {Efremova}, B., {et~al.} 2011, \apss, 335, 161

\bibitem[{{Breeveld} {et~al.}(2010){Breeveld}, {Curran}, {Hoversten}, {Koch},
  {Landsman}, {Marshall}, {Page}, {Poole}, {Roming}, {Smith}, {Still},
  {Yershov}, {Blustin}, {Brown}, {Gronwall}, {Holland}, {Kuin}, {McGowan},
  {Rosen}, {Boyd}, {Broos}, {Carter}, {Chester}, {Hancock}, {Huckle}, {Immler},
  {Ivanushkina}, {Kennedy}, {Mason}, {Morgan}, {Oates}, {de Pasquale},
  {Schady}, {Siegel}, \& {vand en Berk}}]{2010MNRAS.406.1687B}
{Breeveld}, A.~A., {Curran}, P.~A., {Hoversten}, E.~A., {et~al.} 2010, \mnras,
  406, 1687

\bibitem[{{Brown} {et~al.}(2001){Brown}, {Sweigart}, {Lanz}, {Land sman}, \&
  {Hubeny}}]{2001ApJ...562..368B}
{Brown}, T.~M., {Sweigart}, A.~V., {Lanz}, T., {Land sman}, W.~B., \& {Hubeny},
  I. 2001, \apj, 562, 368

\bibitem[{{Brown} {et~al.}(2010){Brown}, {Sweigart}, {Lanz}, {Smith},
  {Landsman}, \& {Hubeny}}]{2010ApJ...718.1332B}
{Brown}, T.~M., {Sweigart}, A.~V., {Lanz}, T., {et~al.} 2010, \apj, 718, 1332

\bibitem[{{Brown} {et~al.}(2016){Brown}, {Cassisi}, {D'Antona}, {Salaris},
  {Milone}, {Dalessandro}, {Piotto}, {Renzini}, {Sweigart}, {Bellini},
  {Ortolani}, {Sarajedini}, {Aparicio}, {Bedin}, {Anderson}, {Pietrinferni}, \&
  {Nardiello}}]{2016ApJ...822...44B}
{Brown}, T.~M., {Cassisi}, S., {D'Antona}, F., {et~al.} 2016, \apj, 822, 44

\bibitem[{Brown {et~al.}(2017)Brown, Taylor, Cassisi, Sweigart, Bellini, Bedin,
  Salaris, Renzini, \& Dalessandro}]{Brown_2017}
Brown, T.~M., Taylor, J.~M., Cassisi, S., {et~al.} 2017, The Astrophysical
  Journal, 851, 118.
\newblock \url{https://doi.org/10.3847%2F1538-4357%2Faa9ce3}

\bibitem[{{Cassisi} \& {Salaris}(2013)}]{2013osp..book.....C}
{Cassisi}, S., \& {Salaris}, M. 2013, {Old Stellar Populations: How to Study
  the Fossil Record of Galaxy Formation}

\bibitem[{{Cassisi} {et~al.}(2008){Cassisi}, {Salaris}, {Pietrinferni},
  {Piotto}, {Milone}, {Bedin}, \& {Anderson}}]{2008ApJ...672L.115C}
{Cassisi}, S., {Salaris}, M., {Pietrinferni}, A., {et~al.} 2008, \apjl, 672,
  L115

\bibitem[{{Castelli} {et~al.}(1997){Castelli}, {Gratton}, \&
  {Kurucz}}]{1997A&A...318..841C}
{Castelli}, F., {Gratton}, R.~G., \& {Kurucz}, R.~L. 1997, \aap, 318, 841

\bibitem[{{Castelli} \& {Kurucz}(2003)}]{2003IAUS..210P.A20C}
{Castelli}, F., \& {Kurucz}, R.~L. 2003, in IAU Symposium, Vol. 210, Modelling
  of Stellar Atmospheres, ed. N.~{Piskunov}, W.~W. {Weiss}, \& D.~F. {Gray},
  A20

\bibitem[{Castelli \& Kurucz(2004)}]{castelli2004new}
Castelli, F., \& Kurucz, R.~L. 2004, New Grids of ATLAS9 Model Atmospheres, , ,
  arXiv:astro-ph/0405087

\bibitem[{{Cohen} {et~al.}(2015){Cohen}, {Hempel}, {Mauro}, {Geisler},
  {Alonso-Garcia}, \& {Kinemuchi}}]{2015AJ....150..176C}
{Cohen}, R.~E., {Hempel}, M., {Mauro}, F., {et~al.} 2015, \aj, 150, 176

\bibitem[{{Conroy} {et~al.}(2009){Conroy}, {Gunn}, \&
  {White}}]{2009ApJ...699..486C}
{Conroy}, C., {Gunn}, J.~E., \& {White}, M. 2009, \apj, 699, 486

\bibitem[{{Cummings} {et~al.}(2014){Cummings}, {Geisler}, {Villanova}, \&
  {Carraro}}]{2014AJ....148...27C}
{Cummings}, J.~D., {Geisler}, D., {Villanova}, S., \& {Carraro}, G. 2014, \aj,
  148, 27

\bibitem[{{Dalessandro} {et~al.}(2008){Dalessandro}, {Lanzoni}, {Ferraro},
  {Rood}, {Milone}, {Piotto}, \& {Valenti}}]{2008ApJ...677.1069D}
{Dalessandro}, E., {Lanzoni}, B., {Ferraro}, F.~R., {et~al.} 2008, \apj, 677,
  1069

\bibitem[{{D'Cruz} {et~al.}(1996){D'Cruz}, {Dorman}, {Rood}, \&
  {O'Connell}}]{1996ApJ...466..359D}
{D'Cruz}, N.~L., {Dorman}, B., {Rood}, R.~T., \& {O'Connell}, R.~W. 1996, \apj,
  466, 359

\bibitem[{{D'Cruz} {et~al.}(2000){D'Cruz}, {O'Connell}, {Rood}, {Whitney},
  {Dorman}, {Landsman}, {Hill}, {Stecher}, \& {Bohlin}}]{2000ApJ...530..352D}
{D'Cruz}, N.~L., {O'Connell}, R.~W., {Rood}, R.~T., {et~al.} 2000, \apj, 530,
  352

\bibitem[{{Dieball} {et~al.}(2009){Dieball}, {Knigge}, {Maccarone}, {Long},
  {Hannikainen}, {Zurek}, \& {Shara}}]{2009MNRAS.394L..56D}
{Dieball}, A., {Knigge}, C., {Maccarone}, T.~J., {et~al.} 2009, \mnras, 394,
  L56

\bibitem[{{Dieball} {et~al.}(2017){Dieball}, {Rasekh}, {Knigge}, {Shara}, \&
  {Zurek}}]{2017MNRAS.469..267D}
{Dieball}, A., {Rasekh}, A., {Knigge}, C., {Shara}, M., \& {Zurek}, D. 2017,
  \mnras, 469, 267

\bibitem[{{Edelmann} {et~al.}(2004){Edelmann}, {Heber}, {Lisker}, \&
  {Green}}]{2004Ap&SS.291..315E}
{Edelmann}, H., {Heber}, U., {Lisker}, T., \& {Green}, E.~M. 2004, \apss, 291,
  315

\bibitem[{{Ferraro} {et~al.}(1999){Ferraro}, {Paltrinieri}, \&
  {Cacciari}}]{1999MmSAI..70..599F}
{Ferraro}, F.~R., {Paltrinieri}, B., \& {Cacciari}, C. 1999, \memsai, 70, 599

\bibitem[{{Ferraro} {et~al.}(1998){Ferraro}, {Paltrinieri}, {Fusi Pecci},
  {Rood}, \& {Dorman}}]{1998ApJ...500..311F}
{Ferraro}, F.~R., {Paltrinieri}, B., {Fusi Pecci}, F., {Rood}, R.~T., \&
  {Dorman}, B. 1998, \apj, 500, 311

\bibitem[{{Gaia Collaboration} {et~al.}(2016){Gaia Collaboration}, {Prusti},
  {de Bruijne}, {Brown}, {Vallenari}, {Babusiaux}, {Bailer-Jones}, {Bastian},
  {Biermann}, {Evans}, {Eyer}, {Jansen}, {Jordi}, {Klioner}, {Lammers},
  {Lindegren}, {Luri}, {Mignard}, {Milligan}, {Panem}, {Poinsignon},
  {Pourbaix}, {Randich}, {Sarri}, {Sartoretti}, {Siddiqui}, {Soubiran},
  {Valette}, {van Leeuwen}, {Walton}, {Aerts}, {Arenou}, {Cropper}, {Drimmel},
  {H{\o}g}, {Katz}, {Lattanzi}, {O'Mullane}, {Grebel}, {Holland}, {Huc},
  {Passot}, {Bramante}, {Cacciari}, {Casta{\~n}eda}, {Chaoul}, {Cheek}, {De
  Angeli}, {Fabricius}, {Guerra}, {Hern{\'a}ndez}, {Jean-Antoine-Piccolo},
  {Masana}, {Messineo}, {Mowlavi}, {Nienartowicz}, {Ord{\'o}{\~n}ez-Blanco},
  {Panuzzo}, {Portell}, {Richards}, {Riello}, {Seabroke}, {Tanga},
  {Th{\'e}venin}, {Torra}, {Els}, {Gracia-Abril}, {Comoretto},
  {Garcia-Reinaldos}, {Lock}, {Mercier}, {Altmann}, {Andrae}, {Astraatmadja},
  {Bellas-Velidis}, {Benson}, {Berthier}, {Blomme}, {Busso}, {Carry},
  {Cellino}, {Clementini}, {Cowell}, {Creevey}, {Cuypers}, {Davidson}, {De
  Ridder}, {de Torres}, {Delchambre}, {Dell'Oro}, {Ducourant}, {Fr{\'e}mat},
  {Garc{\'\i}a-Torres}, {Gosset}, {Halbwachs}, {Hambly}, {Harrison}, {Hauser},
  {Hestroffer}, {Hodgkin}, {Huckle}, {Hutton}, {Jasniewicz}, {Jordan},
  {Kontizas}, {Korn}, {Lanzafame}, {Manteiga}, {Moitinho}, {Muinonen},
  {Osinde}, {Pancino}, {Pauwels}, {Petit}, {Recio-Blanco}, {Robin}, {Sarro},
  {Siopis}, {Smith}, {Smith}, {Sozzetti}, {Thuillot}, {van Reeven}, {Viala},
  {Abbas}, {Abreu Aramburu}, {Accart}, {Aguado}, {Allan}, {Allasia},
  {Altavilla}, {{\'A}lvarez}, {Alves}, {Anderson}, {Andrei}, {Anglada Varela},
  {Antiche}, {Antoja}, {Ant{\'o}n}, {Arcay}, {Atzei}, {Ayache}, {Bach},
  {Baker}, {Balaguer-N{\'u}{\~n}ez}, {Barache}, {Barata}, {Barbier}, {Barblan},
  {Baroni}, {Barrado y Navascu{\'e}s}, {Barros}, {Barstow}, {Becciani},
  {Bellazzini}, {Bellei}, {Bello Garc{\'\i}a}, {Belokurov}, {Bendjoya},
  {Berihuete}, {Bianchi}, {Bienaym{\'e}}, {Billebaud}, {Blagorodnova},
  {Blanco-Cuaresma}, {Boch}, {Bombrun}, {Borrachero}, {Bouquillon}, {Bourda},
  {Bouy}, {Bragaglia}, {Breddels}, {Brouillet}, {Br{\"u}semeister},
  {Bucciarelli}, {Budnik}, {Burgess}, {Burgon}, {Burlacu}, {Busonero}, {Buzzi},
  {Caffau}, {Cambras}, {Campbell}, {Cancelliere}, {Cantat-Gaudin}, {Carlucci},
  {Carrasco}, {Castellani}, {Charlot}, {Charnas}, {Charvet}, {Chassat},
  {Chiavassa}, {Clotet}, {Cocozza}, {Collins}, {Collins}, {Costigan}, {Crifo},
  {Cross}, {Crosta}, {Crowley}, {Dafonte}, {Damerdji}, {Dapergolas}, {David},
  {David}, {De Cat}, {de Felice}, {de Laverny}, {De Luise}, {De March}, {de
  Martino}, {de Souza}, {Debosscher}, {del Pozo}, {Delbo}, {Delgado},
  {Delgado}, {di Marco}, {Di Matteo}, {Diakite}, {Distefano}, {Dolding}, {Dos
  Anjos}, {Drazinos}, {Dur{\'a}n}, {Dzigan}, {Ecale}, {Edvardsson}, {Enke},
  {Erdmann}, {Escolar}, {Espina}, {Evans}, {Eynard Bontemps}, {Fabre},
  {Fabrizio}, {Faigler}, {Falc{\~a}o}, {Farr{\`a}s Casas}, {Faye}, {Federici},
  {Fedorets}, {Fern{\'a}ndez-Hern{\'a}ndez}, {Fernique}, {Fienga}, {Figueras},
  {Filippi}, {Findeisen}, {Fonti}, {Fouesneau}, {Fraile}, {Fraser}, {Fuchs},
  {Furnell}, {Gai}, {Galleti}, {Galluccio}, {Garabato}, {Garc{\'\i}a-Sedano},
  {Gar{\'e}}, {Garofalo}, {Garralda}, {Gavras}, {Gerssen}, {Geyer}, {Gilmore},
  {Girona}, {Giuffrida}, {Gomes}, {Gonz{\'a}lez-Marcos},
  {Gonz{\'a}lez-N{\'u}{\~n}ez}, {Gonz{\'a}lez-Vidal}, {Granvik}, {Guerrier},
  {Guillout}, {Guiraud}, {G{\'u}rpide}, {Guti{\'e}rrez-S{\'a}nchez}, {Guy},
  {Haigron}, {Hatzidimitriou}, {Haywood}, {Heiter}, {Helmi}, {Hobbs},
  {Hofmann}, {Holl}, {Holland }, {Hunt}, {Hypki}, {Icardi}, {Irwin}, {Jevardat
  de Fombelle}, {Jofr{\'e}}, {Jonker}, {Jorissen}, {Julbe}, {Karampelas},
  {Kochoska}, {Kohley}, {Kolenberg}, {Kontizas}, {Koposov}, {Kordopatis},
  {Koubsky}, {Kowalczyk}, {Krone-Martins}, {Kudryashova}, {Kull}, {Bachchan},
  {Lacoste-Seris}, {Lanza}, {Lavigne}, {Le Poncin-Lafitte}, {Lebreton},
  {Lebzelter}, {Leccia}, {Leclerc}, {Lecoeur-Taibi}, {Lemaitre}, {Lenhardt},
  {Leroux}, {Liao}, {Licata}, {Lindstr{\o}m}, {Lister}, {Livanou}, {Lobel},
  {L{\"o}ffler}, {L{\'o}pez}, {Lopez-Lozano}, {Lorenz}, {Loureiro},
  {MacDonald}, {Magalh{\~a}es Fernandes}, {Managau}, {Mann}, {Mantelet},
  {Marchal}, {Marchant}, {Marconi}, {Marie}, {Marinoni}, {Marrese},
  {Marschalk{\'o}}, {Marshall}, {Mart{\'\i}n-Fleitas}, {Martino}, {Mary},
  {Matijevi{\v{c}}}, {Mazeh}, {McMillan}, {Messina}, {Mestre}, {Michalik},
  {Millar}, {Miranda}, {Molina}, {Molinaro}, {Molinaro}, {Moln{\'a}r},
  {Moniez}, {Montegriffo}, {Monteiro}, {Mor}, {Mora}, {Morbidelli}, {Morel},
  {Morgenthaler}, {Morley}, {Morris}, {Mulone}, {Muraveva}, {Musella},
  {Narbonne}, {Nelemans}, {Nicastro}, {Noval}, {Ord{\'e}novic},
  {Ordieres-Mer{\'e}}, {Osborne}, {Pagani}, {Pagano}, {Pailler}, {Palacin},
  {Palaversa}, {Parsons}, {Paulsen}, {Pecoraro}, {Pedrosa}, {Pentik{\"a}inen},
  {Pereira}, {Pichon}, {Piersimoni}, {Pineau}, {Plachy}, {Plum}, {Poujoulet},
  {Pr{\v{s}}a}, {Pulone}, {Ragaini}, {Rago}, {Rambaux}, {Ramos-Lerate},
  {Ranalli}, {Rauw}, {Read}, {Regibo}, {Renk}, {Reyl{\'e}}, {Ribeiro},
  {Rimoldini}, {Ripepi}, {Riva}, {Rixon}, {Roelens}, {Romero-G{\'o}mez},
  {Rowell}, {Royer}, {Rudolph}, {Ruiz-Dern}, {Sadowski}, {Sagrist{\`a}
  Sell{\'e}s}, {Sahlmann}, {Salgado}, {Salguero}, {Sarasso}, {Savietto},
  {Schnorhk}, {Schultheis}, {Sciacca}, {Segol}, {Segovia}, {Segransan},
  {Serpell}, {Shih}, {Smareglia}, {Smart}, {Smith}, {Solano}, {Solitro},
  {Sordo}, {Soria Nieto}, {Souchay}, {Spagna}, {Spoto}, {Stampa}, {Steele},
  {Steidelm{\"u}ller}, {Stephenson}, {Stoev}, {Suess}, {S{\"u}veges}, {Surdej},
  {Szabados}, {Szegedi-Elek}, {Tapiador}, {Taris}, {Tauran}, {Taylor},
  {Teixeira}, {Terrett}, {Tingley}, {Trager}, {Turon}, {Ulla}, {Utrilla},
  {Valentini}, {van Elteren}, {Van Hemelryck}, {van Leeuwen}, {Varadi},
  {Vecchiato}, {Veljanoski}, {Via}, {Vicente}, {Vogt}, {Voss}, {Votruba},
  {Voutsinas}, {Walmsley}, {Weiler}, {Weingrill}, {Werner}, {Wevers},
  {Whitehead}, {Wyrzykowski}, {Yoldas}, {{\v{Z}}erjal}, {Zucker}, {Zurbach},
  {Zwitter}, {Alecu}, {Allen}, {Allende Prieto}, {Amorim},
  {Anglada-Escud{\'e}}, {Arsenijevic}, {Azaz}, {Balm}, {Beck}, {Bernstein},
  {Bigot}, {Bijaoui}, {Blasco}, {Bonfigli}, {Bono}, {Boudreault}, {Bressan},
  {Brown}, {Brunet}, {Bunclark}, {Buonanno}, {Butkevich}, {Carret}, {Carrion},
  {Chemin}, {Ch{\'e}reau}, {Corcione}, {Darmigny}, {de Boer}, {de Teodoro}, {de
  Zeeuw}, {Delle Luche}, {Domingues}, {Dubath}, {Fodor}, {Fr{\'e}zouls},
  {Fries}, {Fustes}, {Fyfe}, {Gallardo}, {Gallegos}, {Gardiol}, {Gebran},
  {Gomboc}, {G{\'o}mez}, {Grux}, {Gueguen}, {Heyrovsky}, {Hoar}, {Iannicola},
  {Isasi Parache}, {Janotto}, {Joliet}, {Jonckheere}, {Keil}, {Kim},
  {Klagyivik}, {Klar}, {Knude}, {Kochukhov}, {Kolka}, {Kos}, {Kutka}, {Lainey},
  {LeBouquin}, {Liu}, {Loreggia}, {Makarov}, {Marseille}, {Martayan},
  {Martinez-Rubi}, {Massart}, {Meynadier}, {Mignot}, {Munari}, {Nguyen},
  {Nordlander}, {Ocvirk}, {O'Flaherty}, {Olias Sanz}, {Ortiz}, {Osorio},
  {Oszkiewicz}, {Ouzounis}, {Palmer}, {Park}, {Pasquato}, {Peltzer}, {Peralta},
  {P{\'e}turaud}, {Pieniluoma}, {Pigozzi}, {Poels}, {Prat}, {Prod'homme},
  {Raison}, {Rebordao}, {Risquez}, {Rocca-Volmerange}, {Rosen}, {Ruiz-Fuertes},
  {Russo}, {Sembay}, {Serraller Vizcaino}, {Short}, {Siebert}, {Silva},
  {Sinachopoulos}, {Slezak}, {Soffel}, {Sosnowska}, {Strai{\v{z}}ys}, {ter
  Linden}, {Terrell}, {Theil}, {Tiede}, {Troisi}, {Tsalmantza}, {Tur},
  {Vaccari}, {Vachier}, {Valles}, {Van Hamme}, {Veltz}, {Virtanen}, {Wallut},
  {Wichmann}, {Wilkinson}, {Ziaeepour}, \& {Zschocke}}]{2016A&A...595A...1G}
{Gaia Collaboration}, {Prusti}, T., {de Bruijne}, J.~H.~J., {et~al.} 2016,
  \aap, 595, A1

\bibitem[{{Gaia Collaboration} {et~al.}(2018{\natexlab{a}}){Gaia
  Collaboration}, {Brown}, {Vallenari}, {Prusti}, {de Bruijne}, {Babusiaux},
  {Bailer-Jones}, {Biermann}, {Evans}, {Eyer}, {Jansen}, {Jordi}, {Klioner},
  {Lammers}, {Lindegren}, {Luri}, {Mignard}, {Panem}, {Pourbaix}, {Randich},
  {Sartoretti}, {Siddiqui}, {Soubiran}, {van Leeuwen}, {Walton}, {Arenou},
  {Bastian}, {Cropper}, {Drimmel}, {Katz}, {Lattanzi}, {Bakker}, {Cacciari},
  {Casta{\~n}eda}, {Chaoul}, {Cheek}, {De Angeli}, {Fabricius}, {Guerra},
  {Holl}, {Masana}, {Messineo}, {Mowlavi}, {Nienartowicz}, {Panuzzo},
  {Portell}, {Riello}, {Seabroke}, {Tanga}, {Th{\'e}venin}, {Gracia-Abril},
  {Comoretto}, {Garcia-Reinaldos}, {Teyssier}, {Altmann}, {Andrae}, {Audard},
  {Bellas-Velidis}, {Benson}, {Berthier}, {Blomme}, {Burgess}, {Busso},
  {Carry}, {Cellino}, {Clementini}, {Clotet}, {Creevey}, {Davidson}, {De
  Ridder}, {Delchambre}, {Dell'Oro}, {Ducourant},
  {Fern{\'a}ndez-Hern{\'a}ndez}, {Fouesneau}, {Fr{\'e}mat}, {Galluccio},
  {Garc{\'\i}a-Torres}, {Gonz{\'a}lez-N{\'u}{\~n}ez}, {Gonz{\'a}lez-Vidal},
  {Gosset}, {Guy}, {Halbwachs}, {Hambly}, {Harrison}, {Hern{\'a}ndez},
  {Hestroffer}, {Hodgkin}, {Hutton}, {Jasniewicz}, {Jean-Antoine-Piccolo},
  {Jordan}, {Korn}, {Krone-Martins}, {Lanzafame}, {Lebzelter}, {L{\"o}ffler},
  {Manteiga}, {Marrese}, {Mart{\'\i}n-Fleitas}, {Moitinho}, {Mora}, {Muinonen},
  {Osinde}, {Pancino}, {Pauwels}, {Petit}, {Recio-Blanco}, {Richards},
  {Rimoldini}, {Robin}, {Sarro}, {Siopis}, {Smith}, {Sozzetti}, {S{\"u}veges},
  {Torra}, {van Reeven}, {Abbas}, {Abreu Aramburu}, {Accart}, {Aerts},
  {Altavilla}, {{\'A}lvarez}, {Alvarez}, {Alves}, {Anderson}, {Andrei},
  {Anglada Varela}, {Antiche}, {Antoja}, {Arcay}, {Astraatmadja}, {Bach},
  {Baker}, {Balaguer-N{\'u}{\~n}ez}, {Balm}, {Barache}, {Barata}, {Barbato},
  {Barblan}, {Barklem}, {Barrado}, {Barros}, {Barstow}, {Bartholom{\'e}
  Mu{\~n}oz}, {Bassilana}, {Becciani}, {Bellazzini}, {Berihuete}, {Bertone},
  {Bianchi}, {Bienaym{\'e}}, {Blanco-Cuaresma}, {Boch}, {Boeche}, {Bombrun},
  {Borrachero}, {Bossini}, {Bouquillon}, {Bourda}, {Bragaglia}, {Bramante},
  {Breddels}, {Bressan}, {Brouillet}, {Br{\"u}semeister}, {Brugaletta},
  {Bucciarelli}, {Burlacu}, {Busonero}, {Butkevich}, {Buzzi}, {Caffau},
  {Cancelliere}, {Cannizzaro}, {Cantat-Gaudin}, {Carballo}, {Carlucci},
  {Carrasco}, {Casamiquela}, {Castellani}, {Castro-Ginard}, {Charlot},
  {Chemin}, {Chiavassa}, {Cocozza}, {Costigan}, {Cowell}, {Crifo}, {Crosta},
  {Crowley}, {Cuypers}, {Dafonte}, {Damerdji}, {Dapergolas}, {David}, {David},
  {de Laverny}, {De Luise}, {De March}, {de Martino}, {de Souza}, {de Torres},
  {Debosscher}, {del Pozo}, {Delbo}, {Delgado}, {Delgado}, {Di Matteo},
  {Diakite}, {Diener}, {Distefano}, {Dolding}, {Drazinos}, {Dur{\'a}n},
  {Edvardsson}, {Enke}, {Eriksson}, {Esquej}, {Eynard Bontemps}, {Fabre},
  {Fabrizio}, {Faigler}, {Falc{\~a}o}, {Farr{\`a}s Casas}, {Federici},
  {Fedorets}, {Fernique}, {Figueras}, {Filippi}, {Findeisen}, {Fonti},
  {Fraile}, {Fraser}, {Fr{\'e}zouls}, {Gai}, {Galleti}, {Garabato},
  {Garc{\'\i}a-Sedano}, {Garofalo}, {Garralda}, {Gavel}, {Gavras}, {Gerssen},
  {Geyer}, {Giacobbe}, {Gilmore}, {Girona}, {Giuffrida}, {Glass}, {Gomes},
  {Granvik}, {Gueguen}, {Guerrier}, {Guiraud}, {Guti{\'e}rrez-S{\'a}nchez},
  {Haigron}, {Hatzidimitriou}, {Hauser}, {Haywood}, {Heiter}, {Helmi}, {Heu},
  {Hilger}, {Hobbs}, {Hofmann}, {Holland}, {Huckle}, {Hypki}, {Icardi},
  {Jan{\ss}en}, {Jevardat de Fombelle}, {Jonker}, {Juh{\'a}sz}, {Julbe},
  {Karampelas}, {Kewley}, {Klar}, {Kochoska}, {Kohley}, {Kolenberg},
  {Kontizas}, {Kontizas}, {Koposov}, {Kordopatis}, {Kostrzewa-Rutkowska},
  {Koubsky}, {Lambert}, {Lanza}, {Lasne}, {Lavigne}, {Le Fustec}, {Le
  Poncin-Lafitte}, {Lebreton}, {Leccia}, {Leclerc}, {Lecoeur-Taibi},
  {Lenhardt}, {Leroux}, {Liao}, {Licata}, {Lindstr{\o}m}, {Lister}, {Livanou},
  {Lobel}, {L{\'o}pez}, {Managau}, {Mann}, {Mantelet}, {Marchal}, {Marchant},
  {Marconi}, {Marinoni}, {Marschalk{\'o}}, {Marshall}, {Martino}, {Marton},
  {Mary}, {Massari}, {Matijevi{\v{c}}}, {Mazeh}, {McMillan}, {Messina},
  {Michalik}, {Millar}, {Molina}, {Molinaro}, {Moln{\'a}r}, {Montegriffo},
  {Mor}, {Morbidelli}, {Morel}, {Morris}, {Mulone}, {Muraveva}, {Musella},
  {Nelemans}, {Nicastro}, {Noval}, {O'Mullane}, {Ord{\'e}novic},
  {Ord{\'o}{\~n}ez-Blanco}, {Osborne}, {Pagani}, {Pagano}, {Pailler},
  {Palacin}, {Palaversa}, {Panahi}, {Pawlak}, {Piersimoni}, {Pineau}, {Plachy},
  {Plum}, {Poggio}, {Poujoulet}, {Pr{\v{s}}a}, {Pulone}, {Racero}, {Ragaini},
  {Rambaux}, {Ramos-Lerate}, {Regibo}, {Reyl{\'e}}, {Riclet}, {Ripepi}, {Riva},
  {Rivard}, {Rixon}, {Roegiers}, {Roelens}, {Romero-G{\'o}mez}, {Rowell},
  {Royer}, {Ruiz-Dern}, {Sadowski}, {Sagrist{\`a} Sell{\'e}s}, {Sahlmann},
  {Salgado}, {Salguero}, {Sanna}, {Santana-Ros}, {Sarasso}, {Savietto},
  {Schultheis}, {Sciacca}, {Segol}, {Segovia}, {S{\'e}gransan}, {Shih},
  {Siltala}, {Silva}, {Smart}, {Smith}, {Solano}, {Solitro}, {Sordo}, {Soria
  Nieto}, {Souchay}, {Spagna}, {Spoto}, {Stampa}, {Steele},
  {Steidelm{\"u}ller}, {Stephenson}, {Stoev}, {Suess}, {Surdej}, {Szabados},
  {Szegedi-Elek}, {Tapiador}, {Taris}, {Tauran}, {Taylor}, {Teixeira},
  {Terrett}, {Teyssand ier}, {Thuillot}, {Titarenko}, {Torra Clotet}, {Turon},
  {Ulla}, {Utrilla}, {Uzzi}, {Vaillant}, {Valentini}, {Valette}, {van Elteren},
  {Van Hemelryck}, {van Leeuwen}, {Vaschetto}, {Vecchiato}, {Veljanoski},
  {Viala}, {Vicente}, {Vogt}, {von Essen}, {Voss}, {Votruba}, {Voutsinas},
  {Walmsley}, {Weiler}, {Wertz}, {Wevers}, {Wyrzykowski}, {Yoldas},
  {{\v{Z}}erjal}, {Ziaeepour}, {Zorec}, {Zschocke}, {Zucker}, {Zurbach}, \&
  {Zwitter}}]{2018A&A...616A...1G}
{Gaia Collaboration}, {Brown}, A.~G.~A., {Vallenari}, A., {et~al.}
  2018{\natexlab{a}}, \aap, 616, A1

\bibitem[{{Gaia Collaboration} {et~al.}(2018{\natexlab{b}}){Gaia
  Collaboration}, {Helmi}, {van Leeuwen}, {McMillan}, {Massari}, {Antoja},
  {Robin}, {Lindegren}, {Bastian}, {Arenou}, {Babusiaux}, {Biermann},
  {Breddels}, {Hobbs}, {Jordi}, {Pancino}, {Reyl{\'e}}, {Veljanoski}, {Brown},
  {Vallenari}, {Prusti}, {de Bruijne}, {Bailer-Jones}, {Evans}, {Eyer},
  {Jansen}, {Klioner}, {Lammers}, {Luri}, {Mignard}, {Panem}, {Pourbaix},
  {Randich}, {Sartoretti}, {Siddiqui}, {Soubiran}, {Walton}, {Cropper},
  {Drimmel}, {Katz}, {Lattanzi}, {Bakker}, {Cacciari}, {Casta{\~n}eda},
  {Chaoul}, {Cheek}, {De Angeli}, {Fabricius}, {Guerra}, {Holl}, {Masana},
  {Messineo}, {Mowlavi}, {Nienartowicz}, {Panuzzo}, {Portell}, {Riello},
  {Seabroke}, {Tanga}, {Th{\'e}venin}, {Gracia-Abril}, {Comoretto},
  {Garcia-Reinaldos}, {Teyssier}, {Altmann}, {Andrae}, {Audard},
  {Bellas-Velidis}, {Benson}, {Berthier}, {Blomme}, {Burgess}, {Busso},
  {Carry}, {Cellino}, {Clementini}, {Clotet}, {Creevey}, {Davidson}, {De
  Ridder}, {Delchambre}, {Dell'Oro}, {Ducourant},
  {Fern{\'a}ndez-Hern{\'a}ndez}, {Fouesneau}, {Fr{\'e}mat}, {Galluccio},
  {Garc{\'\i}a-Torres}, {Gonz{\'a}lez-N{\'u}{\~n}ez}, {Gonz{\'a}lez-Vidal},
  {Gosset}, {Guy}, {Halbwachs}, {Hambly}, {Harrison}, {Hern{\'a}ndez},
  {Hestroffer}, {Hodgkin}, {Hutton}, {Jasniewicz}, {Jean-Antoine-Piccolo},
  {Jordan}, {Korn}, {Krone-Martins}, {Lanzafame}, {Lebzelter}, {L{\"o}ffler},
  {Manteiga}, {Marrese}, {Mart{\'\i}n-Fleitas}, {Moitinho}, {Mora}, {Muinonen},
  {Osinde}, {Pauwels}, {Petit}, {Recio-Blanco}, {Richards}, {Rimoldini},
  {Sarro}, {Siopis}, {Smith}, {Sozzetti}, {S{\"u}veges}, {Torra}, {van Reeven},
  {Abbas}, {Abreu Aramburu}, {Accart}, {Aerts}, {Altavilla}, {{\'A}lvarez},
  {Alvarez}, {Alves}, {Anderson}, {Andrei}, {Anglada Varela}, {Antiche},
  {Arcay}, {Astraatmadja}, {Bach}, {Baker}, {Balaguer-N{\'u}{\~n}ez}, {Balm},
  {Barache}, {Barata}, {Barbato}, {Barblan}, {Barklem}, {Barrado}, {Barros},
  {Barstow}, {Bartholom{\'e} Mu{\~n}oz}, {Bassilana}, {Becciani}, {Bellazzini},
  {Berihuete}, {Bertone}, {Bianchi}, {Bienaym{\'e}}, {Blanco-Cuaresma}, {Boch},
  {Boeche}, {Bombrun}, {Borrachero}, {Bossini}, {Bouquillon}, {Bourda},
  {Bragaglia}, {Bramante}, {Bressan}, {Brouillet}, {Br{\"u}semeister},
  {Brugaletta}, {Bucciarelli}, {Burlacu}, {Busonero}, {Butkevich}, {Buzzi},
  {Caffau}, {Cancelliere}, {Cannizzaro}, {Cantat-Gaudin}, {Carballo},
  {Carlucci}, {Carrasco}, {Casamiquela}, {Castellani}, {Castro-Ginard},
  {Charlot}, {Chemin}, {Chiavassa}, {Cocozza}, {Costigan}, {Cowell}, {Crifo},
  {Crosta}, {Crowley}, {Cuypers}, {Dafonte}, {Damerdji}, {Dapergolas}, {David},
  {David}, {de Laverny}, {De Luise}, {De March}, {de Martino}, {de Souza}, {de
  Torres}, {Debosscher}, {del Pozo}, {Delbo}, {Delgado}, {Delgado}, {Di
  Matteo}, {Diakite}, {Diener}, {Distefano}, {Dolding}, {Drazinos},
  {Dur{\'a}n}, {Edvardsson}, {Enke}, {Eriksson}, {Esquej}, {Eynard Bontemps},
  {Fabre}, {Fabrizio}, {Faigler}, {Falc{\~a}o}, {Farr{\`a}s Casas}, {Federici},
  {Fedorets}, {Fernique}, {Figueras}, {Filippi}, {Findeisen}, {Fonti},
  {Fraile}, {Fraser}, {Fr{\'e}zouls}, {Gai}, {Galleti}, {Garabato},
  {Garc{\'\i}a-Sedano}, {Garofalo}, {Garralda}, {Gavel}, {Gavras}, {Gerssen},
  {Geyer}, {Giacobbe}, {Gilmore}, {Girona}, {Giuffrida}, {Glass}, {Gomes},
  {Granvik}, {Gueguen}, {Guerrier}, {Guiraud}, {Guti{\'e}rrez-S{\'a}nchez},
  {Hofmann}, {Holland}, {Huckle}, {Hypki}, {Icardi}, {Jan{\ss}en}, {Jevardat de
  Fombelle}, {Jonker}, {Juh{\'a}sz}, {Julbe}, {Karampelas}, {Kewley}, {Klar},
  {Kochoska}, {Kohley}, {Kolenberg}, {Kontizas}, {Kontizas}, {Koposov},
  {Kordopatis}, {Kostrzewa-Rutkowska}, {Koubsky}, {Lambert}, {Lanza}, {Lasne},
  {Lavigne}, {Le Fustec}, {Le Poncin-Lafitte}, {Lebreton}, {Leccia}, {Leclerc},
  {Lecoeur-Taibi}, {Lenhardt}, {Leroux}, {Liao}, {Licata}, {Lindstr{\o}m},
  {Lister}, {Livanou}, {Lobel}, {L{\'o}pez}, {Managau}, {Mann}, {Mantelet},
  {Marchal}, {Marchant}, {Marconi}, {Marinoni}, {Marschalk{\'o}}, {Marshall},
  {Martino}, {Marton}, {Mary}, {Matijevi{\v{c}}}, {Mazeh}, {Messina},
  {Michalik}, {Millar}, {Molina}, {Molinaro}, {Moln{\'a}r}, {Montegriffo},
  {Mor}, {Morbidelli}, {Morel}, {Morris}, {Mulone}, {Muraveva}, {Musella},
  {Nelemans}, {Nicastro}, {Noval}, {O'Mullane}, {Ord{\'e}novic},
  {Ord{\'o}{\~n}ez-Blanco}, {Osborne}, {Pagani}, {Pagano}, {Pailler},
  {Palacin}, {Palaversa}, {Panahi}, {Pawlak}, {Piersimoni}, {Pineau}, {Plachy},
  {Plum}, {Poggio}, {Poujoulet}, {Pr{\v{s}}a}, {Pulone}, {Racero}, {Ragaini},
  {Rambaux}, {Ramos-Lerate}, {Regibo}, {Riclet}, {Ripepi}, {Riva}, {Rivard},
  {Rixon}, {Roegiers}, {Roelens}, {Romero-G{\'o}mez}, {Rowell}, {Royer},
  {Ruiz-Dern}, {Sadowski}, {Sagrist{\`a} Sell{\'e}s}, {Sahlmann}, {Salgado},
  {Salguero}, {Sanna}, {Santana-Ros}, {Sarasso}, {Savietto}, {Schultheis},
  {Sciacca}, {Segol}, {Segovia}, {S{\'e}gransan}, {Shih}, {Siltala}, {Silva},
  {Smart}, {Smith}, {Solano}, {Solitro}, {Sordo}, {Soria Nieto}, {Souchay},
  {Spagna}, {Spoto}, {Stampa}, {Steele}, {Steidelm{\"u}ller}, {Stephenson},
  {Stoev}, {Suess}, {Surdej}, {Szabados}, {Szegedi-Elek}, {Tapiador}, {Taris},
  {Tauran}, {Taylor}, {Teixeira}, {Terrett}, {Teyssand ier}, {Thuillot},
  {Titarenko}, {Torra Clotet}, {Turon}, {Ulla}, {Utrilla}, {Uzzi}, {Vaillant},
  {Valentini}, {Valette}, {van Elteren}, {Van Hemelryck}, {van Leeuwen},
  {Vaschetto}, {Vecchiato}, {Viala}, {Vicente}, {Vogt}, {von Essen}, {Voss},
  {Votruba}, {Voutsinas}, {Walmsley}, {Weiler}, {Wertz}, {Wevems},
  {Wyrzykowski}, {Yoldas}, {{\v{Z}}erjal}, {Ziaeepour}, {Zorec}, {Zschocke},
  {Zucker}, {Zurbach}, \& {Zwitter}}]{2018A&A...616A..12G}
{Gaia Collaboration}, {Helmi}, A., {van Leeuwen}, F., {et~al.}
  2018{\natexlab{b}}, \aap, 616, A12

\bibitem[{{Gratton} {et~al.}(2012){Gratton}, {Lucatello}, {Carretta},
  {Bragaglia}, {D'Orazi}, {Al Momany}, {Sollima}, {Salaris}, \&
  {Cassisi}}]{2012A&A...539A..19G}
{Gratton}, R.~G., {Lucatello}, S., {Carretta}, E., {et~al.} 2012, \aap, 539,
  A19

\bibitem[{{Greggio} \& {Renzini}(1990)}]{1990ApJ...364...35G}
{Greggio}, L., \& {Renzini}, A. 1990, \apj, 364, 35

\bibitem[{{Grundahl} {et~al.}(1999){Grundahl}, {Catelan}, {Landsman},
  {Stetson}, \& {Andersen}}]{1999ApJ...524..242G}
{Grundahl}, F., {Catelan}, M., {Landsman}, W.~B., {Stetson}, P.~B., \&
  {Andersen}, M.~I. 1999, \apj, 524, 242

\bibitem[{{Grundahl} {et~al.}(1998){Grundahl}, {VandenBerg}, \&
  {Andersen}}]{1998ApJ...500L.179G}
{Grundahl}, F., {VandenBerg}, D.~A., \& {Andersen}, M.~I. 1998, \apjl, 500,
  L179

\bibitem[{{Harris}(1996)}]{1996AJ....112.1487H}
{Harris}, W.~E. 1996, \aj, 112, 1487

\bibitem[{Hidalgo {et~al.}(2018)Hidalgo, Pietrinferni, Cassisi, Salaris,
  Mucciarelli, Savino, Aparicio, Aguirre, \& Verma}]{Hidalgo_2018}
Hidalgo, S.~L., Pietrinferni, A., Cassisi, S., {et~al.} 2018, The Astrophysical
  Journal, 856, 125.
\newblock \url{https://doi.org/10.3847%2F1538-4357%2Faab158}

\bibitem[{{Hills} \& {Day}(1976)}]{1976ApL....17...87H}
{Hills}, J.~G., \& {Day}, C.~A. 1976, \aplett, 17, 87

\bibitem[{{Jadhav} {et~al.}(2019){Jadhav}, {Sindhu}, \&
  {Subramaniam}}]{2019ApJ...886...13J}
{Jadhav}, V.~V., {Sindhu}, N., \& {Subramaniam}, A. 2019, \apj, 886, 13

\bibitem[{Kim {et~al.}(2019)Kim, Lu, Konopacky, Chu, Toller, Anderson,
  Theissen, \& Morris}]{Kim_2019}
Kim, D., Lu, J.~R., Konopacky, Q., {et~al.} 2019, The Astronomical Journal,
  157, 109.
\newblock \url{https://doi.org/10.3847%2F1538-3881%2Faafb09}

\bibitem[{{Kunder} {et~al.}(2013){Kunder}, {Salaris}, {Cassisi}, {De Propris},
  {Walker}, {Stetson}, {Catelan}, \& {Amigo}}]{2013AJ....145...25K}
{Kunder}, A., {Salaris}, M., {Cassisi}, S., {et~al.} 2013, \aj, 145, 25

\bibitem[{{Leonard}(1989)}]{1989AJ.....98..217L}
{Leonard}, P.~J.~T. 1989, \aj, 98, 217

\bibitem[{{Leone} \& {Manfre}(1997)}]{1997A&A...320..257L}
{Leone}, F., \& {Manfre}, M. 1997, \aap, 320, 257

\bibitem[{{Maxted} {et~al.}(2001){Maxted}, {Heber}, {Marsh}, \&
  {North}}]{2001MNRAS.326.1391M}
{Maxted}, P.~F.~L., {Heber}, U., {Marsh}, T.~R., \& {North}, R.~C. 2001,
  \mnras, 326, 1391

\bibitem[{{McCrea}(1964)}]{1964MNRAS.128..147M}
{McCrea}, W.~H. 1964, \mnras, 128, 147

\bibitem[{{Michaud} {et~al.}(2007){Michaud}, {Richer}, \&
  {Richard}}]{2007ApJ...670.1178M}
{Michaud}, G., {Richer}, J., \& {Richard}, O. 2007, \apj, 670, 1178

\bibitem[{{Milone} {et~al.}(2012){Milone}, {Piotto}, {Bedin}, {Aparicio},
  {Anderson}, {Sarajedini}, {Marino}, {Moretti}, {Davies}, {Chaboyer},
  {Dotter}, {Hempel}, {Mar{\'\i}n-Franch}, {Majewski}, {Paust}, {Reid},
  {Rosenberg}, \& {Siegel}}]{2012A&A...540A..16M}
{Milone}, A.~P., {Piotto}, G., {Bedin}, L.~R., {et~al.} 2012, \aap, 540, A16

\bibitem[{{Milone} {et~al.}(2013){Milone}, {Marino}, {Piotto}, {Bedin},
  {Anderson}, {Aparicio}, {Bellini}, {Cassisi}, {D'Antona}, {Grundahl},
  {Monelli}, \& {Yong}}]{2013ApJ...767..120M}
{Milone}, A.~P., {Marino}, A.~F., {Piotto}, G., {et~al.} 2013, \apj, 767, 120

\bibitem[{{Moehler}(2001)}]{2001PASP..113.1162M}
{Moehler}, S. 2001, \pasp, 113, 1162

\bibitem[{{Moehler} {et~al.}(1995){Moehler}, {Heber}, \& {de
  Boer}}]{1995A&A...294...65M}
{Moehler}, S., {Heber}, U., \& {de Boer}, K.~S. 1995, \aap, 294, 65

\bibitem[{{Moehler} {et~al.}(2019){Moehler}, {Landsman}, {Lanz}, \& {Miller
  Bertolami}}]{2019A&A...627A..34M}
{Moehler}, S., {Landsman}, W.~B., {Lanz}, T., \& {Miller Bertolami}, M.~M.
  2019, \aap, 627, A34

\bibitem[{{Moehler} {et~al.}(2000){Moehler}, {Sweigart}, {Landsman}, \&
  {Heber}}]{2000A&A...360..120M}
{Moehler}, S., {Sweigart}, A.~V., {Landsman}, W.~B., \& {Heber}, U. 2000, \aap,
  360, 120

\bibitem[{Moehler {et~al.}(2000)Moehler, Sweigart, Landsman, \&
  Heber}]{moehler2000hot}
Moehler, S., Sweigart, A.~V., Landsman, W.~B., \& Heber, U. 2000, Hot HB stars
  in globular clusters - Physical parameters and consequences for theory. V.
  Radiative levitation versus helium mixing, , , arXiv:astro-ph/0006182

\bibitem[{{Moehler} {et~al.}(1999){Moehler}, {Sweigart}, {Landsman}, {Heber},
  \& {Catelan}}]{1999A&A...346L...1M}
{Moehler}, S., {Sweigart}, A.~V., {Landsman}, W.~B., {Heber}, U., \& {Catelan},
  M. 1999, \aap, 346, L1

\bibitem[{{Momany} {et~al.}(2004){Momany}, {Bedin}, {Cassisi}, {Piotto},
  {Ortolani}, {Recio-Blanco}, {De Angeli}, \& {Castelli}}]{2004A&A...420..605M}
{Momany}, Y., {Bedin}, L.~R., {Cassisi}, S., {et~al.} 2004, \aap, 420, 605

\bibitem[{{Momany} {et~al.}(2002){Momany}, {Piotto}, {Recio-Blanco}, {Bedin},
  {Cassisi}, \& {Bono}}]{2002ApJ...576L..65M}
{Momany}, Y., {Piotto}, G., {Recio-Blanco}, A., {et~al.} 2002, \apjl, 576, L65

\bibitem[{{Momany} {et~al.}(2020){Momany}, {Zaggia}, {Montalto}, {Jones},
  {Boffin}, {Cassisi}, {Moni Bidin}, {Gullieuszik}, {Saviane}, {Monaco},
  {Mason}, {Girardi}, {D'Orazi}, {Piotto}, {Milone}, {Lala}, {Stetson}, \&
  {Beletsky}}]{2020NatAs.tmp..116M}
{Momany}, Y., {Zaggia}, S., {Montalto}, M., {et~al.} 2020, Nature Astronomy,
  arXiv:2006.02308

\bibitem[{{Moni Bidin}(2018)}]{2018OAst...27...91M}
{Moni Bidin}, C. 2018, Open Astronomy, 27, 91

\bibitem[{{Moni Bidin} {et~al.}(2009){Moni Bidin}, {Moehler}, {Piotto},
  {Momany}, \& {Recio-Blanco}}]{2009A&A...498..737M}
{Moni Bidin}, C., {Moehler}, S., {Piotto}, G., {Momany}, Y., \& {Recio-Blanco},
  A. 2009, \aap, 498, 737

\bibitem[{{Moni Bidin, C.} {et~al.}(2006){Moni Bidin, C.}, {Moehler, S.},
  {Piotto, G.}, {Recio-Blanco, A.}, {Momany, Y.}, \& {M\'endez, R.
  A.}}]{refId0}
{Moni Bidin, C.}, {Moehler, S.}, {Piotto, G.}, {et~al.} 2006, A\&A, 451, 499.
\newblock \url{https://doi.org/10.1051/0004-6361:20053940}

\bibitem[{{Napiwotzki} {et~al.}(2004){Napiwotzki}, {Karl}, {Lisker}, {Heber},
  {Christlieb}, {Reimers}, {Nelemans}, \& {Homeier}}]{2004Ap&SS.291..321N}
{Napiwotzki}, R., {Karl}, C.~A., {Lisker}, T., {et~al.} 2004, \apss, 291, 321

\bibitem[{{Nardiello} {et~al.}(2018){Nardiello}, {Libralato}, {Piotto},
  {Anderson}, {Bellini}, {Aparicio}, {Bedin}, {Cassisi}, {Granata}, {King},
  {Lucertini}, {Marino}, {Milone}, {Ortolani}, {Platais}, \& {van der
  Marel}}]{2018MNRAS.481.3382N}
{Nardiello}, D., {Libralato}, M., {Piotto}, G., {et~al.} 2018, \mnras, 481,
  3382

\bibitem[{{Pace} {et~al.}(2006){Pace}, {Recio-Blanco}, {Piotto}, \&
  {Momany}}]{2006A&A...452..493P}
{Pace}, G., {Recio-Blanco}, A., {Piotto}, G., \& {Momany}, Y. 2006, \aap, 452,
  493

\bibitem[{{Page} {et~al.}(2012){Page}, {Brindle}, {Talavera}, {Still}, {Rosen},
  {Yershov}, {Ziaeepour}, {Mason}, {Cropper}, {Breeveld}, {Loiseau}, {Mignani},
  {Smith}, \& {Murdin}}]{2012MNRAS.426..903P}
{Page}, M.~J., {Brindle}, C., {Talavera}, A., {et~al.} 2012, \mnras, 426, 903

\bibitem[{{Piotto} {et~al.}(2015){Piotto}, {Milone}, {Bedin}, {Anderson},
  {King}, {Marino}, {Nardiello}, {Aparicio}, {Barbuy}, {Bellini}, {Brown},
  {Cassisi}, {Cool}, {Cunial}, {Dalessandro}, {D'Antona}, {Ferraro}, {Hidalgo},
  {Lanzoni}, {Monelli}, {Ortolani}, {Renzini}, {Salaris}, {Sarajedini}, {van
  der Marel}, {Vesperini}, \& {Zoccali}}]{2015AJ....149...91P}
{Piotto}, G., {Milone}, A.~P., {Bedin}, L.~R., {et~al.} 2015, \aj, 149, 91

\bibitem[{Poole {et~al.}(2007)Poole, Breeveld, Page, Landsman, Holland, Roming,
  Kuin, Brown, Gronwall, Hunsberger, Koch, Mason, Schady, Berk, Blustin, Boyd,
  Broos, Carter, Chester, Cucchiara, Hancock, Huckle, Immler, Ivanushkina,
  Kennedy, Marshall, Morgan, Pandey, De~Pasquale, Smith, \&
  Still}]{10.1111/j.1365-2966.2007.12563.x}
Poole, T.~S., Breeveld, A.~A., Page, M.~J., {et~al.} 2007, Monthly Notices of
  the Royal Astronomical Society, 383, 627.
\newblock \url{https://doi.org/10.1111/j.1365-2966.2007.12563.x}

\bibitem[{{Roming} {et~al.}(2005){Roming}, {Kennedy}, {Mason}, {Nousek}, {Ahr},
  {Bingham}, {Broos}, {Carter}, {Hancock}, {Huckle}, {Hunsberger}, {Kawakami},
  {Killough}, {Koch}, {McLelland}, {Smith}, {Smith}, {Soto}, {Boyd},
  {Breeveld}, {Holland}, {Ivanushkina}, {Pryzby}, {Still}, \&
  {Stock}}]{2005SSRv..120...95R}
{Roming}, P. W.~A., {Kennedy}, T.~E., {Mason}, K.~O., {et~al.} 2005, \ssr, 120,
  95

\bibitem[{{Rood} {et~al.}(1998){Rood}, {Dorman}, {Ferraro}, {Paltrinieri}, \&
  {Fusi Pecci}}]{1998ESASP.413..515R}
{Rood}, R.~T., {Dorman}, B., {Ferraro}, F.~R., {Paltrinieri}, B., \& {Fusi
  Pecci}, F. 1998, in ESA Special Publication, Vol. 413, Ultraviolet
  Astrophysics Beyond the IUE Final Archive, ed. W.~{Wamsteker}, R.~{Gonzalez
  Riestra}, \& B.~{Harris}, 515

\bibitem[{Saffer {et~al.}(1998)Saffer, Livio, \& Yungelson}]{Saffer_1998}
Saffer, R.~A., Livio, M., \& Yungelson, L.~R. 1998, The Astrophysical Journal,
  502, 394.
\newblock \url{https://doi.org/10.1086%2F305907}

\bibitem[{{Sahu} {et~al.}(2019{\natexlab{a}}){Sahu}, {Subramaniam},
  {C{\^o}t{\'e}}, {Rao}, \& {Stetson}}]{2019MNRAS.482.1080S}
{Sahu}, S., {Subramaniam}, A., {C{\^o}t{\'e}}, P., {Rao}, N.~K., \& {Stetson},
  P.~B. 2019{\natexlab{a}}, \mnras, 482, 1080

\bibitem[{{Sahu} {et~al.}(2019{\natexlab{b}}){Sahu}, {Subramaniam},
  {Simunovic}, {Postma}, {C{\^o}t{\'e}}, {Kameswera Rao}, {Geller}, {Leigh},
  {Shara}, {Puzia}, \& {Stetson}}]{2019ApJ...876...34S}
{Sahu}, S., {Subramaniam}, A., {Simunovic}, M., {et~al.} 2019{\natexlab{b}},
  \apj, 876, 34

\bibitem[{{Sandage}(1953)}]{1953AJ.....58...61S}
{Sandage}, A.~R. 1953, \aj, 58, 61

\bibitem[{{Sanders}(1971)}]{1971A&A....14..226S}
{Sanders}, W.~L. 1971, \aap, 14, 226

\bibitem[{{Siegel} {et~al.}(2014){Siegel}, {Porterfield}, {Linevsky}, {Bond},
  {Holland }, {Hoversten}, {Berrier}, {Breeveld}, {Brown}, \&
  {Gronwall}}]{2014AJ....148..131S}
{Siegel}, M.~H., {Porterfield}, B.~L., {Linevsky}, J.~S., {et~al.} 2014, \aj,
  148, 131

\bibitem[{{Sindhu} {et~al.}(2019){Sindhu}, {Subramaniam}, {Jadhav},
  {Chatterjee}, {Geller}, {Knigge}, {Leigh}, {Puzia}, {Shara}, \&
  {Simunovic}}]{2019ApJ...882...43S}
{Sindhu}, N., {Subramaniam}, A., {Jadhav}, V.~V., {et~al.} 2019, \apj, 882, 43

\bibitem[{Singh \& Yadav(2018)}]{10.1093/mnras/sty2961}
Singh, G., \& Yadav, R. K.~S. 2018, Monthly Notices of the Royal Astronomical
  Society, 482, 4874.
\newblock \url{https://doi.org/10.1093/mnras/sty2961}

\bibitem[{{Sirianni} {et~al.}(2005){Sirianni}, {Jee}, {Ben{\'\i}tez},
  {Blakeslee}, {Martel}, {Meurer}, {Clampin}, {De Marchi}, {Ford}, {Gilliland
  }, {Hartig}, {Illingworth}, {Mack}, \& {McCann}}]{2005PASP..117.1049S}
{Sirianni}, M., {Jee}, M.~J., {Ben{\'\i}tez}, N., {et~al.} 2005, \pasp, 117,
  1049

\bibitem[{{Sosin} {et~al.}(1997){Sosin}, {Dorman}, {Djorgovski}, {Piotto},
  {Rich}, {King}, {Liebert}, {Phinney}, \& {Renzini}}]{1997ApJ...480L..35S}
{Sosin}, C., {Dorman}, B., {Djorgovski}, S.~G., {et~al.} 1997, \apjl, 480, L35

\bibitem[{{Stetson}(1987)}]{1987PASP...99..191S}
{Stetson}, P.~B. 1987, \pasp, 99, 191

\bibitem[{{Stetson} {et~al.}(2019){Stetson}, {Pancino}, {Zocchi}, {Sanna}, \&
  {Monelli}}]{2019MNRAS.485.3042S}
{Stetson}, P.~B., {Pancino}, E., {Zocchi}, A., {Sanna}, N., \& {Monelli}, M.
  2019, \mnras, 485, 3042

\bibitem[{{Subramaniam} {et~al.}(2016{\natexlab{a}}){Subramaniam}, {Sindhu},
  {Tandon}, {Kameswara Rao}, {Postma}, {C{\^o}t{\'e}}, {Hutchings}, {Ghosh},
  {George}, {Girish}, {Mohan}, {Murthy}, {Sankarasubramanian}, {Stalin},
  {Sutaria}, {Mondal}, \& {Sahu}}]{2016ApJ...833L..27S}
{Subramaniam}, A., {Sindhu}, N., {Tandon}, S.~N., {et~al.} 2016{\natexlab{a}},
  \apjl, 833, L27

\bibitem[{{Subramaniam} {et~al.}(2016{\natexlab{b}}){Subramaniam}, {Tandon},
  {Hutchings}, {Ghosh}, {George}, {Girish}, {Kamath}, {Kathiravan}, {Kumar},
  {Lancelot}, {Mahesh}, {Mohan}, {Murthy}, {Nagabhushana}, {Pati}, {Postma},
  {Rao}, {Sankarasubramanian}, {Sreekumar}, {Sriram}, {Stalin}, {Sutaria},
  {Sreedhar}, {Barve}, {Mondal}, \& {Sahu}}]{2016SPIE.9905E..1FS}
{Subramaniam}, A., {Tandon}, S.~N., {Hutchings}, J., {et~al.}
  2016{\natexlab{b}}, Society of Photo-Optical Instrumentation Engineers (SPIE)
  Conference Series, Vol. 9905, {In-orbit performance of UVIT on ASTROSAT},
  99051F

\bibitem[{{Subramaniam} {et~al.}(2017){Subramaniam}, {Sahu}, {Postma},
  {C{\^o}t{\'e}}, {Hutchings}, {Darukhanawalla}, {Chung}, {Tandon}, {Kameswara
  Rao}, {George}, {Ghosh}, {Girish}, {Mohan}, {Murthy}, {Pati},
  {Sankarasubramanian}, {Stalin}, \& {Choudhury}}]{2017AJ....154..233S}
{Subramaniam}, A., {Sahu}, S., {Postma}, J.~E., {et~al.} 2017, \aj, 154, 233

\bibitem[{{Sweigart}(1997)}]{1997ApJ...474L..23S}
{Sweigart}, A.~V. 1997, \apjl, 474, L23

\bibitem[{{Sweigart}(2002)}]{2002HiA....12..292S}
---. 2002, Highlights of Astronomy, 12, 292

\bibitem[{{Tandon} {et~al.}(2017){Tandon}, {Subramaniam}, {Girish}, {Postma},
  {Sankarasubramanian}, {Sriram}, {Stalin}, {Mondal}, {Sahu}, {Joseph},
  {Hutchings}, {Ghosh}, {Barve}, {George}, {Kamath}, {Kathiravan}, {Kumar},
  {Lancelot}, {Leahy}, {Mahesh}, {Mohan}, {Nagabhushana}, {Pati}, {Kameswara
  Rao}, {Sreedhar}, \& {Sreekumar}}]{2017AJ....154..128T}
{Tandon}, S.~N., {Subramaniam}, A., {Girish}, V., {et~al.} 2017, \aj, 154, 128

\bibitem[{{Zinn} \& {Searle}(1976)}]{1976ApJ...209..734Z}
{Zinn}, R., \& {Searle}, L. 1976, \apj, 209, 734

\bibitem[{Zwart \& Leigh(2019)}]{Zwart_2019}
Zwart, S.~P., \& Leigh, N. W.~C. 2019, The Astrophysical Journal, 876, L33.
\newblock \url{https://doi.org/10.3847%2F2041-8213%2Fab1b75}

\end{thebibliography}

\end{document}